\documentclass[journal,twoside]{IEEEtran}

\usepackage{amsmath,amsfonts}
\usepackage{algorithmic}
\usepackage{algorithm}
\usepackage{array}
\usepackage[caption=false,font=normalsize,labelfont=sf,textfont=sf]{subfig}
\usepackage{textcomp}
\usepackage{stfloats}
\usepackage{url}
\usepackage{verbatim}
\usepackage{graphicx}

\usepackage{cite}
\usepackage{placeins} 

\usepackage{color,array,amsthm}
\usepackage{bm} 
\usepackage{mathabx}

\usepackage{csquotes}
\usepackage{balance}
\usepackage{xcolor}
\usepackage[colorlinks=true, allcolors=blue]{hyperref}

\usepackage{enumitem}  % 再引入该宏包

\def\BibTeX{{\rm B\kern-.05em{\sc i\kern-.025em b}\kern-.08em
    T\kern-.1667em\lower.7ex\hbox{E}\kern-.125emX}}

\begin{document}

\title{Deep-Learning-Based Classification of Relay-Surface Scattering Regimes for NLOS Millimeter-Wave Sensing}

\author{Haoran Yuan,
        Guanghua Liu,~\IEEEmembership{Senior Member,~IEEE,}

        You Xu,~\IEEEmembership{Member,~IEEE,}
        Zhihao Liu

        % <-this % stops a space
% <-this % stops a space
\thanks{Corresponding author: Guanghua Liu.}
\thanks{The authors are with Underground Networking and Intelligent Communications Laboratory, School of Cyber Science and Engineering,  Huazhong University of Science and Technology, Wuhan 430074, China (e-mail: haoranyuan@hust.edu.cn; guanghualiu@hust.edu.cn; 
youxu@ieee.org; liuzhihao6@hust.edu.cn).}
}

\markboth{IEEE Journal on Multiscale and Multiphysics Computational Techniques}
{Yuan \MakeLowercase{\textit{et al.}}: Deep-Learning-Based Classification of Relay-Surface Scattering Regimes}

\maketitle

{\fontsize{7.5}{9}\selectfont\noindent\textit{This work has been submitted to the IEEE for possible publication. Copyright may be transferred without notice, after which this version may no longer be accessible.}\par}
\vspace{0.5\baselineskip}

\begin{abstract}
In non-line-of-sight (NLOS) millimeter-wave sensing, relay-surface scattering alters target echoes and can cause propagation-model mismatch, angle-estimation errors, and ghost targets. DEKL, a deep-learning framework, identifies the dominant scattering regime from radar echoes without measuring surface roughness. Multipath echo models describe specular reflection, structured multipath reflection, and diffuse scattering. These models and a no-target hypothesis define a model-selection problem with unknown parameters. The Kullback--Leibler information criterion yields an adaptive decision rule for selecting among these hypotheses. A lightweight network, KLNet, learns covariance-domain features and combines attention with confidence-gated iterative decoding to avoid repeated likelihood evaluation. Analytical simulations, independent electromagnetic simulations, and radar measurements show that DEKL improves discrimination over the evaluated information-criterion-based rules. Ablations show how attention, confidence supervision, and gated decoding contribute to selective prediction and robustness under controlled covariance perturbations. The predicted regime guides propagation-model selection for radio-environment representations in wireless-network digital twins and informs NLOS localization and imaging.
\end{abstract}


\section{INTRODUCTION}
W{\scshape ireless} sensing provides observations of propagation environments, linking localization and environmental awareness to wireless-network digital twins. Such twins require computational representations of radio propagation, and neural scene models support wireless-channel prediction and network planning \cite{cao2024raypronet,bakirtzis2025rigorous}. At millimeter-wave frequencies, non-line-of-sight (NLOS) sensing exploits relay-surface reflections to observe hidden targets \cite{pham2024around}. Representative third-order-bounce geometries connect a sensing node to an occluded target \cite{danino2024automatic,xu2023leveraging}, supporting environmental monitoring and mobility applications \cite{xu2024double,zheng2024detection}. However, scattering may produce unequal directions of departure (DODs) and arrival (DOAs) \cite{xu2023leveraging,levy2023mcrb,xu2025bayesian}. Model mismatch then complicates echo interpretation \cite{milczarek2023automatic}, motivating identification of the dominant propagation mechanism before localization.

NLOS sensing methods have addressed multipath propagation through signal modeling and reconstruction \cite{xu2023leveraging,zhu2023non}. Early propagation-identification frameworks distinguish direct and indirect paths \cite{borras1998decision}. For colocated radar arrays, matched departure and arrival angles characterize the simplest return geometry, whereas additional reflections introduce angular discrepancies. Optical NLOS sensing and specular-scattering analysis also illustrate the importance of modeling illumination-surface interactions \cite{buttafava2015non,fan2024specular}, although their measurement configurations differ from microwave arrays. Rough-surface microstructure introduces distributed scattering contributions whose statistical properties affect detection and wireless propagation \cite{rong2020diffuse,jiang2022survey}. Unlike a single specular component, these returns need not concentrate around one angular pair. Specular and diffuse components may coexist within the received echo \cite{aubry2015diffuse}, producing multipath clusters with varying angular and delay characteristics. Modeling scattering above rough surfaces further requires accounting for interactions between objects and their surroundings \cite{wei2021rough}. Uncertainty in path strengths and angular structure complicates fixed steering-vector descriptions. Multipath detection methods consequently exploit spatially distributed observations \cite{lian2024distributed}.

Relay-surface scattering depends on roughness as well as observation geometry and electromagnetic surface properties. Measuring these properties for every surface in a complex scene is generally impractical. Millimeter-wave sensing studies have therefore examined multipath discrimination \cite{zheng2024detection} and wall-reflection characteristics for polarimetric NLOS identification \cite{alburadi2025rapid}. When path parameters are uncertain, joint DOD and DOA estimation can recover coupled angular information \cite{wen2019joint}, while multipath returns can contribute useful detection information \cite{lian2024distributed}. DOD--DOA relationships offer observable cues for multipath recognition \cite{luo2023effective}, complemented by measurements of wave-interacting objects \cite{guzman2023wave}. For relay-assisted sensing, the question is which dominant scattering model best represents the observed echoes. This calls for a decision from sensing features rather than direct inversion of surface roughness, with unknown propagation parameters treated within candidate models. The selected category describes the received signal under the assumed geometry, rather than an intrinsic material label.

Computational efficiency and physical fidelity are central to propagation modeling for wireless applications. Learning-based indoor propagation models incorporate environmental and antenna information to accelerate wireless-system simulations \cite{bakirtzis2025rigorous}. These forward predictors motivate efficient environmental representations, whereas the present task selects a dominant echo model from observations. In NLOS sensing, hierarchical sparse priors represent intergroup sparsity among multipath components and improve reconstruction \cite{xu2024double}, while multipath consistency supports simultaneous LOS and NLOS localization with clutter mitigation \cite{chen2025joint}. Sparse Bayesian learning (SBL) offers a framework for joint DOD and DOA estimation \cite{cao2019sparse}. Modeling the coupling between departure and arrival angles helps resolve localization ambiguities associated with DOD\(\neq\)DOA \cite{zhu2023non}, but higher-order reflections and diffuse returns complicate parameter estimation. These approaches address reconstruction or localization within their respective propagation descriptions. Adaptive detection under multiple alternatives likewise involves fitting and comparing candidate models \cite{addabbo2021adaptive2}. As the number of unknown paths grows, parameter searches and likelihood evaluations become demanding. Reduced signal sparsity further challenges sparse estimation \cite{ge20205g}. Identifying the dominant echo regime and assessing prediction reliability before applying a specialized processing model is therefore a distinct task. This distinction matters when multiple scattering mechanisms coexist within the same resolution cell.

In this paper, we develop an adaptive framework for identifying dominant relay-surface scattering regimes in NLOS millimeter-wave wireless sensing, using multiple-input multiple-output (MIMO) radar echoes as observations. The framework selects representative propagation models with unknown scattering attributes and assumed environmental geometry. Specular reflection, structured multipath reflection, diffuse scattering, and a no-target hypothesis define the candidate set. The multiple-hypothesis formulation links echo structure to statistical model selection. The computational objective is to select an appropriate echo representation efficiently when the underlying propagation parameters are unknown and repeated likelihood evaluations become costly. Information-criterion-based model-order selection \cite{carotenuto2017model} supplies the decision formulation, while learned signal features \cite{walenczykowska2022application} enable efficient classification. The resulting DEKL framework uses a lightweight network, KLNet, to extract covariance-domain features and produce classification and confidence scores. Its design is inspired by the Kullback--Leibler information criterion (KLIC). Analytical simulations, independent electromagnetic simulations, and measured data assess scattering discrimination. The resulting observation-driven model selection supports propagation modeling in radio-environment and wireless-network digital twins by linking sensing observations to computational representations of the physical environment.

The main contributions of this paper are summarized as follows.
\begin{enumerate}[topsep=0pt]
  \item We formulate relay-surface scattering identification as a multiple-hypothesis model-selection problem for NLOS wireless sensing. Specular, structured multipath, and diffuse echo models connect observable structure to candidate propagation descriptions with unknown parameters, without directly recovering surface roughness.

  \item We develop a KLIC-inspired adaptive decision framework and implement its classification stage with a lightweight neural network. Covariance-domain feature extraction, attention-based feature weighting, and confidence supervision are combined with shared-weight iterative decoding, whose stopping rule uses class stability and confidence. This enables hypothesis selection without repeated likelihood evaluation or exhaustive parameter searches.

  \item We evaluate the framework using analytical simulations for controlled training, independent finite-difference time-domain (FDTD) simulations for testing under physical model mismatch, and independent measured millimeter-wave radar data. Comparisons demonstrate improved discrimination over the considered criteria, while ablations establish the contributions of attention, confidence supervision, and gated decoding to selective prediction and robustness under controlled covariance perturbations.
\end{enumerate}

\section{SIGNAL MODELS AND PROBLEM FORMULATION}

\begin{figure}[!t]
\centerline{\includegraphics[width=22pc]{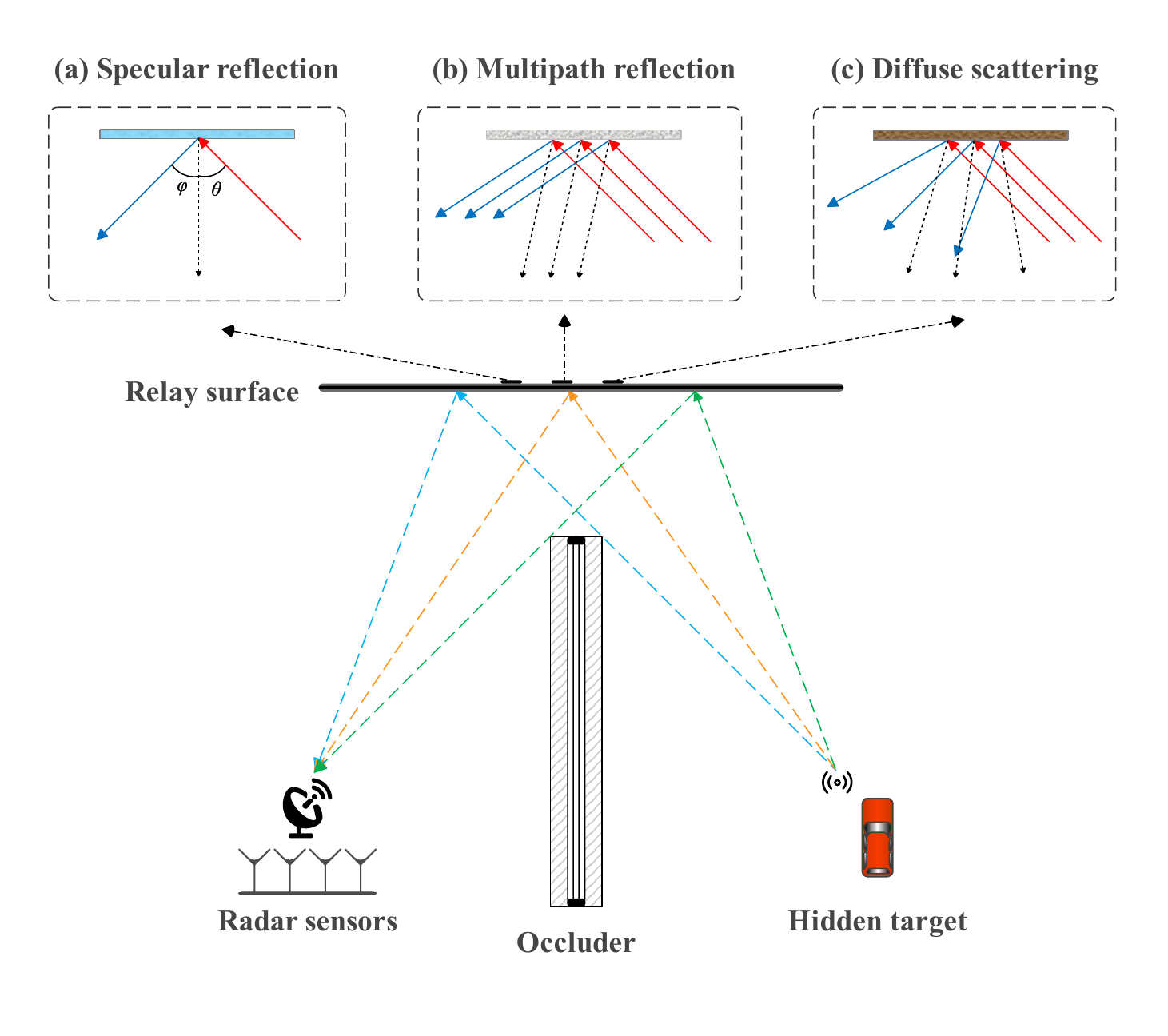}}
\caption{NLOS scenarios of relay-surface-reflected radar signals.}
\label{fig1}
\end{figure}

Consider an NLOS scenario depicted in Fig. \ref{fig1}, where a vehicle is the detection target. It is assumed that the environmental layout, including the position and geometry of relay surfaces, is known a priori \cite{xu2023leveraging}. This information, obtained through methods such as the fusion of Lidar and millimeter-wave radar data for relay-surface estimation and signal filtering \cite{danino2024automatic}, is considered prerequisite knowledge and is not the focus of this study. As illustrated in the figure, an impenetrable, nonreflective obstacle wall obstructs the direct signal path, so no valid echo component is generated by either transmission through or reflection from the occluder. Consequently, all useful NLOS target returns are assumed to arise from relay-surface-assisted propagation, whose reflection characteristics vary with the physical properties of the specific relay surface involved. Identifying their dominant scattering supports propagation-model selection for wireless-network digital twins.

As established in the literature \cite{ben2008surface}, when a millimeter wave impinges on a relay surface at an incident angle $\theta_{i}$ measured from the surface normal, increasing roughness generally favors diffuse scattering at a fixed wavelength and incidence geometry. The surface unevenness height, denoted as $\Delta h$, provides the following approximate roughness criteria.
\[\begin{aligned}\Delta h&<\frac{\lambda}{25\cos\theta_i}&\text{Smooth surface}\\\frac{\lambda}{25\cos\theta_i}&<\Delta h<\frac{\lambda}{8\cos\theta_i}&\text{Moderately rough surface}\\\Delta h&>\frac{\lambda}{8\cos\theta_i}&\text{Rough surface}\end{aligned}\]

Since surface roughness in real-world environments exhibits continuous, multiscale characteristics, echo signals typically manifest the physical properties of mixed-mode scattering \cite{nazari2022random}. To reduce computational complexity and enable rapid selection of signal processing strategies, a common engineering approximation is to discretize these characteristics into dominant scattering mechanisms \cite{rafique2025rough}. Accordingly, an approximate correspondence can be established between the roughness level of the relay surface and the reflection type of the echo signal under the assumed geometry and material conditions \cite{ma2019terahertz}.

For a smooth surface satisfying the criterion, specular reflection typically dominates. Under the assumed colocated geometry with a retraced propagation path, DOD = DOA. This manifests as a diagonal concentration in the joint DOD--DOA angular spectrum, and the corresponding echo typically exhibits low delay spread. Its signal-to-noise ratio (SNR) also depends on path loss and target reflectivity. For a moderately rough surface, multipath reflection can produce distinct DOD and DOA (DOD \(\neq\) DOA). In the paired-path model, this manifests as reciprocal angular pairs offset from the diagonal in the DOD--DOA domain, with the echo characterized by multipath delay spread and phase fluctuations. Finally, a rough surface can produce predominantly diffuse scattering, with an irregular distribution of dense clusters in the DOD--DOA domain. The resulting echo exhibits anisotropic scattering intensity and random fluctuations. Therefore, instead of directly estimating the continuous physical roughness parameters of the relay surface, this work reformulates the recognition of the dominant scattering mechanism from the observed echo features as a multivariate hypothesis testing problem.

At the receiver, a colocated MIMO radar is employed. The system comprises $M_{T}$ transmit antennas and $M_{R}$ receive antennas, yielding $N=M_TM_R$ transmit--receive channels after matched filtering. By collecting measurements, we construct a data matrix  $Z=[z_1,...,z_K]\in\mathbb{C}^{N\times K}$, whose columns are modeled as statistically independent random vectors  \cite{zheng2024detection}. It is reasonable to assume that these vectors share the same distributional parameters, and their jointly unknown probability density function (PDF) is denoted by $\bar{f}(Z;\theta)$. Here, $\theta\in\mathbb{R}^{\mathrm{p}\times1}$ is a parameter vector taking values in a parameter space $\Theta\subseteq\mathbb{R}^{\mathrm{p}\times1}$. Let each transmit antenna emit \(L\) pulses, and let $x(l)=\begin{bmatrix}x_1(l),x_2(l),\cdotp\cdotp\cdotp,x_{M_T}(l)\end{bmatrix}^T$ denote the code vector transmitted by the $M_T$ antennas at the $\mathit{l}$-th time instant. The transmit code matrix is then given by $X=[x(1),x(2),\cdots,x(L)]\in\mathbb{C}^{M_T\times L}$. After applying a Fast Fourier Transform (FFT) along the fast-time dimension of the received measurements, the multipath scenarios can be modeled separately under the various hypotheses.

\begin{itemize}
    \item Case I: Noise Only
\end{itemize}

Under this hypothesis ($H_0$), the received signal consists solely of noise processed by the matched filter, as modeled by
\begin{equation}
\mathbf{r}=\left(\left(X^*P(f_d)\right)\otimes I_{M_R}\right)\widetilde{w}. \label{eq:1}
\end{equation}
In this equation, $\widetilde{w} = \mathrm{vec}(W)$ is the vectorized form of the noise matrix $W=[w(1),w(2),\cdots,w(L)]$. Each column $w(l) \in \mathbb{C}^{M_R \times 1}$ represents a complex white Gaussian noise vector at the receiving array, following the distribution $\mathcal{CN}(0, \sigma^2 I_{M_R})$, where $\sigma^2$ is the noise variance. The term $f_d$ denotes the normalized Doppler frequency, and $P(f_d)$ is a diagonal matrix defined as
\begin{equation}
P(f_d)=\mathrm{diag}\left(\left[1, e^{j2\pi f_d}, \cdots, e^{j2\pi f_d(L-1)}\right]\right). \label{eq:2}
\end{equation}

\begin{itemize}
    \item Case II: Specular Reflection
\end{itemize}

Under the specular reflection hypothesis ($H_1$), the received signal $\mathbf{z}$ is modeled as the sum of $K_0$ effective direct path components and noise:
\begin{equation}
\mathbf{z}=\left(\mathbf{R}_x\otimes\boldsymbol{I}_{M_R}\right)\sum_{k=1}^{K_0}\alpha_k\mathbf{a}_T(\theta_k)\otimes\mathbf{a}_R(\theta_k)+\mathbf{r}.
\label{eq:3}
\end{equation}
Here, $\mathbf{R}_x = X^* X^T$ is the transmit correlation matrix. The vector $\mathbf{\Theta}_0 = [\theta_1, \theta_2, \ldots, \theta_{K_0}]^T \in \mathbb{R}^{K_0 \times 1}$ contains the angles of the $K_0$ direct paths. $\theta_k$ denotes DOD = DOA of the $\mathit{k}$-th direct path, and the corresponding steering matrix is given by
\begin{equation}
\mathbf{A}(\mathbf{\Theta}_{0})=\left[\mathbf{a}(\mathbf{\theta}_{1}),\mathbf{a}(\mathbf{\theta}_{2}),\ldots,\mathbf{a}(\mathbf{\theta}_{K_{0}})\right]\in\mathbb{C}^{M_{T}M_{R}\times K_{0}},
\label{eq:4}
\end{equation}
where the steering vector is defined as $\mathbf{a}(\cdot)= \mathbf{a}_T(\cdot) \otimes \mathbf{a}_R(\cdot)$:
\begin{subequations}\label{eq:5}
\begin{align}
\mathbf{a}_T(\theta)=\frac{1}{\sqrt{M_T}}[&e^{j2\pi d_{T,1}\sin(\theta)/\lambda}, \ldots, e^{j2\pi d_{T,M_T}\sin(\theta)/\lambda}]^T,
\label{eq:a} \\
\mathbf{a}_R(\phi)=\frac{1}{\sqrt{M_R}}[&e^{j2\pi d_{R,1}\sin(\phi)/\lambda}, \ldots, e^{j2\pi d_{R,M_R}\sin(\phi)/\lambda}]^T.
\label{eq:b}
\end{align}
\end{subequations}
In these expressions, $\theta$ and $\phi$ are the departure and arrival angles, respectively; $\lambda$ is the wavelength; and $d_{T,m}$ and $d_{R,n}$ are the positions of the $m$-th transmit and $n$-th receive antenna elements relative to a reference point. Using this notation, the signal model in \eqref{eq:3} can be simplified to
\begin{equation}
{\mathbf{z}}=(\mathbf{R}_x\otimes\boldsymbol{I}_{M_R})\mathbf{A}(\Theta_0)\bm{\alpha}+\mathbf{r},
\label{eq:6}
\end{equation}
where $\bm{\alpha}=\left[\alpha_1,\alpha_2,...,\alpha_{K_0}\right]^T\in\mathbb{C}^{K_0\times1}$ is the complex amplitude vector for the direct paths.

\begin{itemize}
    \item Case III: Multipath Reflection
\end{itemize}

Under the multipath reflection hypothesis ($H_2$), the received signal $\mathbf{z}$ includes components from direct paths, as in $H_1$, as well as $K_1$ distinct multipath components. The signal model is given by
\begin{equation}
\begin{aligned}
\mathbf{z} = & (\mathbf{R}_x\otimes\boldsymbol{I}_{M_R})\sum_{k=1}^{K_0}\alpha_k\mathbf{a}_T(\theta_k)\otimes\mathbf{a}_R(\theta_k) \\
& + (\mathbf{R}_x\otimes\boldsymbol{I}_{M_R})\sum_{k=1}^{K_1}\beta_{k,1}\mathbf{a}_T(\vartheta_k)\otimes\mathbf{a}_R(\varphi_k) \\
& + (\mathbf{R}_x\otimes\boldsymbol{I}_{M_R})\sum_{k=1}^{K_1}\beta_{k,2}\mathbf{a}_T(\varphi_k)\otimes\mathbf{a}_R(\vartheta_k) + \mathbf{r}.
\end{aligned}
\label{eq:7}
\end{equation}
The DODs for the $K_1$ multipath reflection paths are collected in $\mathbf{\Theta}_{1}=[\vartheta_{1},\ldots,\vartheta_{K_{1}}]^{T}$, and the DOAs are collected in $\mathbf{\Phi}_1=[\varphi_1,\ldots,\varphi_{K_1}]^T$; typically, $\vartheta_k \neq \varphi_k$. Define the angle sets $\mathbf{\Theta}= [\mathbf{\Theta}_{1}^{T}, \mathbf{\Phi}_{1}^{T}, \mathbf{\Theta}_{0}^{T}]^{T} \in\mathbb{R}^{(2K_{1}+K_{0})\times1}$ and $\mathbf{\Phi}=[\mathbf{\Phi}_1^T, \mathbf{\Theta}_1^T, \mathbf{\Theta}_0^T]^T \in\mathbb{R}^{(2K_1+K_0)\times1}$. The corresponding transmit and receive steering matrices are then constructed as
\begin{subequations}\label{eq:8}
\begin{align} 
\mathbf{A}_{T}(\mathbf{\Theta}) = & \left[\mathbf{a}_{T}(\vartheta_{1}), \ldots, \mathbf{a}_{T}(\vartheta_{K_{1}}), \mathbf{a}_{T}(\varphi_{1}), \ldots, \mathbf{a}_{T}(\varphi_{K_{1}}), \right. \nonumber\\
& \left. \mathbf{a}_{T}(\theta_{1}), \ldots, \mathbf{a}_{T}(\theta_{K_{0}})\right], \\
\mathbf{A}_{R}(\mathbf{\Phi}) = & \left[\mathbf{a}_{R}(\varphi_{1}), \ldots, \mathbf{a}_{R}(\varphi_{K_{1}}), \mathbf{a}_{R}(\vartheta_{1}), \ldots, \mathbf{a}_{R}(\vartheta_{K_{1}}), \right. \nonumber\\
& \left. \mathbf{a}_{R}(\theta_{1}), \ldots, \mathbf{a}_{R}(\theta_{K_{0}})\right].
\end{align}
\end{subequations}
Similarly, the signal model in \eqref{eq:7} can be compactly expressed as
\begin{equation}
\mathbf{z}=(\mathbf{R}_{x}\otimes\boldsymbol{I}_{M_{R}})\mathbf{A}(\mathbf{\Theta},\mathbf{\Phi})\bm{\beta}+\mathbf{r},
\label{eq:9}
\end{equation}
where $\mathbf{A}(\mathbf{\Theta},\mathbf{\Phi})=\mathbf{A}_T(\mathbf{\Theta})\circ\mathbf{A}_R(\mathbf{\Phi})$ denotes the response matrix, $\bm{\beta}=[{\beta}_1^{T},{\alpha}^{T}]^T \in\mathbb{C}^{(2K_1+K_0)\times1}$ is the composite amplitude vector, and $\beta_1=[\beta_{1,1},\beta_{2,1},\ldots,\beta_{K_1,1},\beta_{1,2},\beta_{2,2},\ldots,\beta_{K_1,2}]^{T}\in\mathbb{C}^{2K_1\times1}$ is the amplitude vector for the $K_1$ multipath pairs.

\begin{itemize}
    \item Case IV: Diffuse Scattering
\end{itemize}

Under the diffuse scattering hypothesis ($H_3$), the scattering surface in each range bin is assumed to generate $N_{ML}$ echoes that impinge within the main lobe of the transmit beam pattern, while $N_{SL}$ echoes originate from sidelobe directions. The aggregate echo from each patch is modeled as a complex, zero-mean, circularly symmetric Gaussian process. These processes are statistically independent and follow a multivariate complex normal distribution with an unknown covariance matrix:
\begin{equation}
\mathbf{z}=(\mathbf{R}_x\otimes\boldsymbol{I}_{M_R})\sum_{i=1}^{N_s}\alpha_i\mathbf{p}+(\mathbf{R}_x\otimes\boldsymbol{I}_{M_R})\mathbf{D}+\mathbf{r},
\label{eq:10}
\end{equation}
where $\alpha_{i}=\alpha\frac{x_{i}}{\sqrt{q_{\mathrm{mp}}}}$, $i=1,\ldots,N_{ML}+N_{SL}$, and $x_i$ are independent, zero-mean, circularly symmetric complex Gaussian variables with unit variance; $\alpha \in \mathbb{C}$ is an unknown deterministic parameter that accounts for target reflectivity and channel propagation effects; and $q_{\mathrm{mp}}>0$ is the multipath power-normalization parameter. The vector $\mathbf{p}=\frac{\mathbf{a}_{T}\otimes \mathbf{a}_{R}}{\|\mathbf{a}_{T}\|_{2}\cdot\|\mathbf{a}_{R}\|_{2}}$ is the normalized steering vector associated with the direct path. $\mathbf{D}\sim\mathcal{CN}(0,\Sigma(\alpha,q_{\mathrm{mp}}))$ is the diffuse multipath matrix, with $\Sigma(\alpha,q_{\mathrm{mp}})$ being its covariance matrix.

Summarizing, the primary data covariance matrix is given by:
\begin{equation}
\mathbf{M}=\sigma_c^2\overline{\mathbf{M}}_1+\sigma_n^2\boldsymbol{I}+\Sigma(\alpha,q_{\mathrm{mp}}),
\label{eq:11}
\end{equation}
where $\overline{\mathbf{M}}_1(n,m)=e^{-(n-m)^2/(2\sigma_d^2)}$ denotes the $(n,m)$-th element of the normalized clutter covariance matrix. The correlation-width parameter is $\sigma_d=0.995$, $\sigma_n^2>0$ is the thermal noise power, and $\sigma_c^2$ is the clutter power. The diffuse multipath covariance is
\begin{equation}
\Sigma(\alpha,q_{\mathrm{mp}})=\sum_{i=1}^{N_{s}}\frac{|\alpha|^2}{q_{\mathrm{mp}}}\mathbf{v}(\theta_i)\mathbf{v}(\theta_i)^\dagger.
\label{eq:12}
\end{equation}

Let $N_{s}=N_{ML}+N_{SL}$, and let $L_{msl}$ denote the multipath straddle loss (additional loss) in the model. The multipath covariance matrix is expressed as:
\begin{equation}
\Sigma(\alpha,q_{\mathrm{mp}})
=\sum_{i=1}^{N_s}
\frac{\sigma}{L_{msl}}
\overline{\mathbf{v}}(\phi_i,\theta_i)
\overline{\mathbf{v}}(\phi_i,\theta_i)^\dagger,
\label{eq:13}
\end{equation}
where $\overline{\mathbf{v}}(\phi, \theta) = \frac{1}{\sqrt{N}} [1, e^{j2\pi \bar{\xi}(\phi, \theta)}, \ldots, e^{j2\pi (N-1) \bar{\xi}(\phi, \theta)}]^T$ and $\bar{\xi}(\phi, \theta) = \frac{d}{\lambda} \cos(\phi) \sin(\theta)$. Here, $N = M_T M_R$, and $\sigma$, as the average scattering intensity coefficient of the rough surface, replaces the undetermined parameter $\alpha$.

Ultimately, the task of classifying the multipath scattering from a relay surface is framed as a multivariate hypothesis testing problem involving these four hypotheses ($H_0, H_1, H_2, H_3$). These models encode angular constraints, path pairing, and diffuse covariance. For mixed-mode observations, the selected hypothesis should be interpreted as the closest dominant mechanism, while the confidence score introduced later is used to indicate ambiguous or transitional cases. As indicated by prior literature \cite{carotenuto2017model,barthelme2020model}, model-order selection (MOS) criterion based on minimizing information distance is particularly effective for selecting the optimal mathematical model to describe observed data, especially in adaptive detection tasks within complex environments.

The multiple-hypothesis decision problem is divided into multiple subsets, corresponding to the null hypothesis ($H_{0}$) and the alternative hypotheses ($H_1, H_2, H_3$). The set of collected echo signals, $Z$, is classified into one of these four categories. Each hypothesis is associated with angle sets $\Theta_i=\{\theta_1,\theta_2,\ldots,\theta_k|k_i\}$ and $\Phi_i=\{\phi_1,\phi_2,\ldots,\phi_k|k_i\}$, a path count $k_{i}$, and other unknown parameters such as a confidence level $p_{n}$. The PDF of $Z$ under the $n$-th hypothesis can be written as $f_{n,i}(Z;\Theta_i,\Phi_i;k_i;p_n)$, for $n=0,1,2,3$. It is crucial to distinguish between the unknown PDF of $Z$, denoted $\bar{f}(Z;\Theta)$, and the actual PDF, $f_i(Z;\Theta_i)$. The observable model order is then $k_m = k_i + k_s$. The decision problem can be summarized as:
\begin{equation}
\begin{cases}
    H_{0}: Z \sim f_{0}(\dots) = r, & \text{Noise} \\
    H_{1}: Z \sim f_{1,i}(\dots) = \alpha_{1}A_{i}(\Theta_{i},\Phi_{i})+r, & \text{$\Theta_i=\Phi_i$} \\
    H_{2}: Z \sim f_{2,i}(\dots) = \alpha_{2}A_{i}(\Theta_{i},\Phi_{i})+r, & \text{$\Theta_i\neq\Phi_i$} \\
    H_{3}: Z \sim f_{3,i}(\dots) = \alpha_{3}A_{i}(\Theta_{i},\Phi_{i})+r, & \text{$\Theta_i\stackrel{?}{=}\Phi_i$}
\end{cases}
\label{eq:14}
\end{equation}

It should be noted that $H_0$--$H_3$ do not imply mutually exclusive pure physical surfaces. They are used as representative statistical models for dominant echo structures, and mixed-scattering cases are handled through confidence-aware decision making.

\section{PROPOSED METHOD}
Define the sample vector as $\vec{\mathbf{y}} = [y_1, y_2,\ldots,y_k]^T$, and the conditional probability as $p_i(\vec{\mathbf{y}})=p(y_1,y_2,\ldots,y_k|H_i), 0\leq i\leq M-1$. The decision is made in the $k$-dimensional space of $\vec{\mathbf{y}}$, which is partitioned into $M$ decision regions, $\mathcal{R}_0, \mathcal{R}_1, \ldots, \mathcal{R}_{M-1}$. If $\vec{\mathbf{y}}$ falls within region $\mathcal{R}_i$, the decision is $H_i$. Under the classical MOS criterion, this decision problem can be formulated as an optimization problem:
\begin{equation}
\hat{L} = \underset{\ell \in \{0, \ldots, L_{\max}\}}{\operatorname{argmax}}
\left[\ln p_\ell(\mathbf{Y}; \hat{\Theta}, \hat{R}_s, \hat{\sigma}_\eta^2)-c(\ell)\right],
\label{eq:15}
\end{equation}
where $\mathbf{Y}$ represents the set of $k$ measurements $y_1, y_2,\ldots,y_k$. The estimated model order is $\hat{L}$, and the model parameters are estimated via maximum-likelihood (ML) estimation. The likelihood function of the received signal is denoted by $p_\ell(\cdot)$, and $c(\ell)\geq0$ is a complexity penalty used to mitigate overfitting induced by excessive model order, following the same subtractive convention as $h(n)$ below.

However, the presence of outliers, clutter edges, and regions with strongly varying reflectivity may invalidate the assumption that the true model belongs to the family of candidate models. Accordingly, a more general MOS rule is pursued: we adopt a Taylor series approximation to the Kullback--Leibler Divergence (KLD), apply an appropriate prior to the unknown model order, and combine ML with Bayesian estimation to regularize the PDF under general alternative hypotheses. Consequently, a model detector based on the Kullback--Leibler Information Criterion (KLIC) is developed.

\subsection{Parameter Estimation}

Under the null hypothesis $H_{0}$, which assumes a pure-noise scenario with no target present, there are no additional unknown parameters. Therefore, parameter estimation is unnecessary. Only the alternative hypotheses of specular reflection, multipath reflection, and diffuse scattering are considered.

\begin{itemize}
    \item $H_{1}$ Hypothesis
\end{itemize}
Under the alternative hypothesis $H_{1}$, we attempt to solve the problem
\begin{equation}
\begin{aligned} &(\hat{K}_0, \widehat{\Theta_0}, \hat{\alpha}) = \underset{K_0, \Theta_0, \alpha}{\operatorname*{\arg\min}} \quad K_0 \\
&\quad \text{s.t.} \quad \left\|\bar{\mathbf{z}}-\bar{\mathbf{A}}(\Theta_0)\alpha\right\|_2^2\leq\epsilon^2,\end{aligned}
\label{eq:16}  % \label{} 用于标记公式
\end{equation}
where $\bar{\mathbf{A}}(\cdot)=\Sigma_x^{1/2}\mathbf{A}(\cdot)$. This problem is solved using an iterative algorithm, which is initialized with the residual $\mathbf{e}^{(0)}=\bar{\mathbf{z}}$. In each iteration $t$, a new path angle is identified from a predefined grid $\{\tilde{\theta}_1, \ldots, \tilde{\theta}_G\}$ by solving
\begin{equation}
\hat{\theta}^{(t)}=\underset{\theta^{(t)}\in\{\tilde{\theta}_1, \ldots, \tilde{\theta}_G\}}{\operatorname*{\arg\max}}\left|\left(\mathbf{e}^{(t-1)}\right)^H\mathbf{a}(\theta^{(t)})\right|.
\label{eq:17}
\end{equation}
This grid-based selection admits a principled interpretation from a greedy sparse approximation viewpoint. Given the residual $\mathbf{e}^{(t-1)}$ and a candidate steering atom $\mathbf{a}(\theta)$ on the grid (here $\theta$ denotes an arbitrary grid point), the least-squares amplitude update is $\hat{\alpha}(\theta)=\frac{\mathbf{a}(\theta)^H\mathbf{e}^{(t-1)}}{\|\mathbf{a}(\theta)\|_2^2}$, which yields
\begin{equation}
\left\|\mathbf{e}^{(t-1)}-\mathbf{a}(\theta)\hat{\alpha}(\theta)\right\|_2^2
=\left\|\mathbf{e}^{(t-1)}\right\|_2^2-\frac{\left|\mathbf{a}(\theta)^H\mathbf{e}^{(t-1)}\right|^2}{\|\mathbf{a}(\theta)\|_2^2}.
\label{eq:18}
\end{equation}
Therefore, maximizing $\left|\left(\mathbf{e}^{(t-1)}\right)^H\mathbf{a}(\theta)\right|$ in \eqref{eq:17} achieves the largest one-step reduction in residual energy among grid candidates (note that $|\mathbf{e}^H\mathbf{a}|=|\mathbf{a}^H\mathbf{e}|$), and provides a reliable coarse initialization for subsequent continuous refinement.
Update the angle matrix as $\widehat{\Theta}_0^{(t,0)}=[(\widehat{\Theta}_0^{(t-1)})^T, \hat{\theta}^{(t)}]^T$. Subsequently, take $\widehat{\Theta}_0^{(t,0)}$ (the grid-search result) as the initial point and apply the Gauss--Newton (GN) iteration to improve the accuracy of the estimate:
\begin{equation}
\widehat{\Theta}_0^{(t,i+1)}=\widehat{\Theta}_0^{(t,i)}-(\mathbf{H}_0^{(t,i)})^{-1}\mathbf{g}_0^{(t,i)},
\label{eq:19}
\end{equation}
where $\mathbf{g}_0^{(t,i)}$ and $\mathbf{H}_{0}^{(t,i)}$ denote, respectively, the gradient and the Gauss--Newton approximation to the Hessian of the cost function $F(\Theta_{0}^{(t,i)}) = \|\bar{\mathbf{z}}-\bar{\mathbf{A}}(\Theta_{0}^{(t,i)})\bar{\mathbf{A}}^{+}(\Theta_{0}^{(t,i)})\bar{\mathbf{z}}\|_{2}^{2}$. Here, $(\cdot)^+$ denotes the Moore--Penrose pseudoinverse, whereas $(\cdot)^\dagger$ denotes the Hermitian transpose. \eqref{eq:19} updates the continuous angle parameters by linearizing the residual at the current estimate and minimizing the resulting quadratic approximation of the nonlinear least-squares objective. The full closed-form expressions for $\mathbf{g}_0^{(t,i)}$ and $\mathbf{H}_{0}^{(t,i)}$ are lengthy and are omitted for brevity.

When the iteration reaches the maximum count $I$, the refined angle estimate obtained in the $t$-th iteration, $\widehat{\Theta_{0}}^{(t,i)}$, is taken as $\widehat{\Theta}_{0}^{(t)}$. The amplitude estimate and the residual are then updated as follows:
\begin{subequations}\label{eq:20}
\begin{align}
\hat{\alpha}^{(t)} &= \bar{\mathbf{A}}^+(\widehat{\Theta}_0^{(t)})\bar{\mathbf{z}}, \\
\mathbf{e}^{(t)} &= \bar{\mathbf{z}}-\bar{\mathbf{A}}(\widehat{\Theta}_0^{(t)})\hat{\alpha}^{(t)}.
\end{align}
\end{subequations}
The iteration terminates once the stopping criterion $\|\mathbf{e}^{(t)}\|_2 \leq \varepsilon_1$ is met.

\begin{itemize}
    \item $H_{2}$ Hypothesis
\end{itemize}
Under hypothesis $H_2$, the estimation problem is formulated as
\begin{equation}
\begin{aligned}
&(\hat{K}_{0}, \hat{K}_{1}, \widehat{\Theta}, \widehat{\Phi}, \hat{\beta}) = \underset{\substack{K_{0}, K_{1}, \Theta, \Phi, \beta}}{\operatorname*{\arg\min}} \quad K_{0} + \delta K_{1} \\
&\qquad \text{s.t.} \quad \|\bar{\mathbf{z}} - \bar{\mathbf{A}}(\Theta, \Phi)\beta\|_{2}^{2} \leq \epsilon^{2},
\end{aligned}
\label{eq:21}
\end{equation}
where $\delta$ is a tuning parameter that controls the relative weight between $K_0$ and $K_1$. The operator $(\Theta,\Phi)$ is evaluated over a $G$-dimensional grid of common DOD--DOA values to obtain coarse angle estimates, assembling the set $\bar{\bar{\Theta}}^{(t,0)}=[\bar{\bar{\Theta}}_{1}^{(t,0)};\bar{\bar{\Phi}}_{1}^{(t,0)};\bar{\bar{\Theta}}_{0}^{(t,0)}]$, which is constructed as
\begin{subequations}\label{eq:22}
\begin{align}
\bar{\bar{\Theta}}_1^{(t,0)} &= [(\widehat{\Theta}_1^{(t-1)})^T, \hat{\vartheta}^{(t)}]^T, \\
\bar{\bar{\Phi}}_1^{(t,0)} &= [(\widehat{\Phi}_1^{(t-1)})^T, \hat{\varphi}^{(t)}]^T, \\
\bar{\bar{\Theta}}_0^{(t,0)} &= \widehat{\Theta}_0^{(t-1)}.
\end{align}
\end{subequations}
The Levenberg--Marquardt (LM) method is employed to refine the angle estimates. We temporarily substitute $\bar{\bar{\Theta}}^{(t,0)}$ for $\bar{\Theta}^{(t,0)}$, and the updated angle set is
\begin{equation}
\bar{\Theta}^{(t,i+1)}=\bar{\Theta}^{(t,i)}-\left(\mathbf{H}^{(t,i)}+\mu^{(t,i)}\boldsymbol{I}_{\widehat{K}}^{(t)}\right)^{-1}\mathrm{g}^{(t,i)},
\label{eq:23}
\end{equation}
where the cost function is given by $\bar{F}(\bar{\Theta}) = \|\bar{\mathbf{z}}-\bar{\mathbf{A}}(\bar{\Theta},\bar{\Phi})\bar{\mathbf{A}}^{+}(\bar{\Theta},\bar{\Phi})\bar{\mathbf{z}}\|_{2}^{2}$. The damping coefficient $\mu^{(t,i)}$ is selected by a gain-ratio-guided line search. This damping improves numerical stability when the local curvature is ill-conditioned by providing a smooth transition between a Gauss--Newton step and a steepest-descent step. Since paths with unequal DOD and DOA are introduced in pairs---i.e., reflected multipath components always occur in paired form with group sparsity---it is necessary to further partition the multipath components.

The angle set $\bar{\Theta}^{(t,I)}=[\bar{\Theta}_{1}^{(t,I)};\bar{\Phi}_{1}^{(t,I)};\bar{\Theta}_{0}^{(t,I)}]$ for estimating the additional specular direct path, together with the corresponding residual $\mathbf{e}_{1}^{(t)}$ defined analogously to \eqref{eq:20}, and the angle set $\bar{\bar{\Theta}}^{(t,I)}=[\bar{\bar{\Theta}}_{1}^{(t,I)};\bar{\bar{\Phi}}_{1}^{(t,I)};\bar{\bar{\Theta}}_{0}^{(t,I)}]$ for estimating the additional multipath component, together with the corresponding residual $\mathbf{e}_{2}^{(t)}$, are obtained. Finally, based on the difference $\mathbf{e}^{(t)}=\|\mathbf{e}_{2}^{(t)}\|_2-\|\mathbf{e}_{1}^{(t)}\|_2$, we determine which angle-set estimate is more consistent with the observations. The specific decision rules are as follows:
\begin{align}
&(\widehat{\Theta}_1^{(t)},\widehat{\Phi}_1^{(t)},\widehat{\Theta}_0^{(t)}) \notag \\
&=\begin{cases}
(\bar{\Theta}_1^{(t,I)},\bar{\Phi}_1^{(t,I)},\bar{\Theta}_0^{(t,I)}) & \mathbf{e}^{(t)}\geq\delta_{r} \\
(\bar{\bar{\Theta}}_1^{(t,I)},\bar{\bar{\Phi}}_1^{(t,I)},\bar{\bar{\Theta}}_0^{(t,I)}) & \mathbf{e}^{(t)}<\delta_{r}
\end{cases}
\label{eq:24}
\end{align}
where $\delta_{r}$ is a properly chosen threshold. At this point, the iterative procedure terminates once the condition $\|\mathbf{e}^{(t-1)}\|_2-\|\mathbf{e}^{(t)}\|_2\leq\varepsilon_{2}$ is satisfied.

\begin{itemize}
    \item $H_{3}$ Hypothesis
\end{itemize}

\begin{figure*}[!t]
\centerline{\includegraphics[width=42pc]{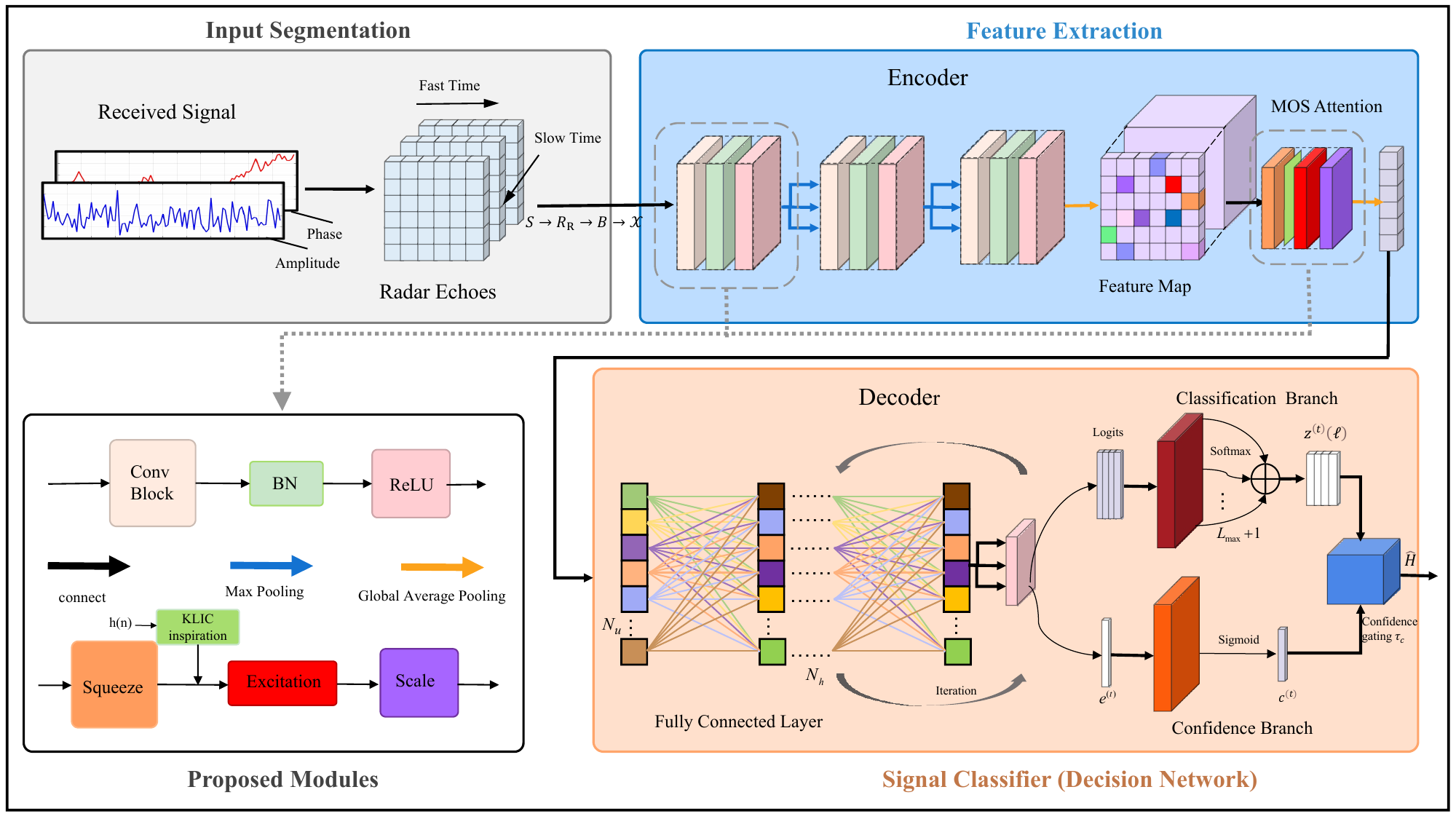}}
\caption{Overall framework of the proposed DEKL method. The encoder extracts hierarchical spatial correlation features from the input beamspace covariance tensor, while the MOS attention module introduces penalty-aware feature weighting. The decoder performs iterative feature refinement and outputs classification probabilities alongside confidence scores for confidence-gated decision making.}
\label{fig2}
\end{figure*}

Under hypothesis $H_3$, the parameters are estimated by solving the following optimization problem:
\begin{equation}
\begin{aligned}&(\widehat{\Theta},\widehat{\Phi},\widehat{\alpha},\widehat{\sigma})=
\underset{\alpha,\{\theta_i,\phi_i\},\sigma}{\operatorname*{\arg\min}}
\biggl[\boldsymbol{T}+\log\det(\mathbf{M})\biggr]\\
&\qquad \boldsymbol{T}=(\mathbf{z}-\alpha\boldsymbol{p})^\dagger\mathbf{M}^{-1}
(\mathbf{z}-\alpha\boldsymbol{p})\\
&\quad \text{s.t.} \quad
\left\|\boldsymbol{S}^{1/2}\mathbf{M}^{-1}\boldsymbol{S}^{1/2}
-\boldsymbol{I}\right\|_2\leq\epsilon, 
\quad \mathbf{M}\succ0,
\end{aligned}
\label{eq:25}
\end{equation}
where $\boldsymbol{S}=\frac{1}{K}\sum_{l=1}^{K}\mathbf{z}_{l}\mathbf{z}_{l}^{\dagger}=\begin{bmatrix}\widehat{m}_{11}&\widehat{m}_{21}^{\dagger}\\\widehat{m}_{21}&\widehat{M}_{22}\end{bmatrix}$ denotes the sample covariance computed from $K$ independent auxiliary samples $\{\mathbf{z}_l\}_{l=1}^K$. Since the angle set of the diffuse scattering path is randomly distributed within a certain range, the neighborhood radius $\epsilon$ must be determined by expected likelihood (EL) estimation of the diffuse scattering covariance $\mathbf{M}$ over the constraint set $\Omega_{\epsilon}$:
\begin{equation}
\hat{\epsilon}_{\mathrm{EL}}=\operatorname{arg\min}_{\epsilon\geq0}|LR(\epsilon|\mathrm{z})-LR_a|^2,
\label{eq:26}
\end{equation}
where $LR_{a}$ is a preset threshold and the likelihood ratio $LR(\epsilon|\mathrm{z})$ is
\begin{equation}
LR(\epsilon|\mathrm{z})=\prod_{h=1}^{H_G}\prod_{i=1}^r[\hat{\lambda}_{h,i}(\epsilon)\hat{\omega}_{h,i}e^{1-\hat{\lambda}_{h,i}(\epsilon)\hat{\omega}_{h,i}}]^{H_L}.
\label{eq:27}
\end{equation}
In this equation, the columns of the diffuse multipath matrix $\mathbf{D}$ are clustered into several groups sharing the same covariance matrix, $H_G$ is the number of data groups, and $H_L$ is the number of samples in each group. $\hat{\lambda}_{h,\mathrm{i}}(\epsilon)=\min\{1+\epsilon,\max(1/\hat{\omega}_{h,i},1-\epsilon)\}$, where $\hat{\omega}_{h,i}$ is the eigenvalue of the submatrix $\frac{1}{H_{L}}\mathrm{Z}_{2h}^{\dagger}\widehat{M}_{22}^{-1}\mathrm{Z}_{2h}$. 

Within the constrained domain radius $\hat{\epsilon}_{\mathrm{EL}}$, the MUSIC algorithm is used to search for angle initialization:
\begin{subequations}\label{eq:28}
\begin{align}
    &(\hat{\phi}_i^{(0)},\hat{\theta}_i^{(0)})=\operatorname{arg\max}_{\phi,\theta}P_{\mathrm{MUSIC}}\left(\phi,\theta\right), \\
    &P_{\mathrm{MUSIC}}(\phi,\theta)=\frac{1}{|\overline{\mathbf{v}}(\phi,\theta)^\dagger \mathbf{U}_n\mathbf{U}_n^\dagger\overline{\mathbf{v}}(\phi,\theta)|},
\end{align}
\end{subequations}
where $\mathbf{U}_{n}$ denotes the noise subspace from the eigendecomposition of the sample covariance $\boldsymbol{S}$. The peak locations of the MUSIC spectrum correspond to the initial angle estimates $(\hat{\theta}_i^{(0)},\hat{\phi}_i^{(0)})$. Initialize the multipath covariance estimate as $\widehat{\mathbf{M}}^{(0)}=\hat{\sigma}^{(0)}\sum_{i=1}^{N_{s}}\overline{\mathbf{v}}\left(\hat{\phi}_{i}^{(0)},\hat{\theta}_{i}^{(0)}\right)\overline{\mathbf{v}}\left(\hat{\phi}_{i}^{(0)},\hat{\theta}_{i}^{(0)}\right)^{\dagger}$, and
\begin{subequations}\label{eq:29}
\begin{align}\hat{\alpha}^{(0)}&=\frac{\mathbf{p}^\dagger\left(\widehat{\mathbf{M}}^{(0)}\right)^{-1}\mathbf{z}}{\mathbf{p}^\dagger\left(\widehat{\mathbf{M}}^{(0)}\right)^{-1}\mathbf{p}}, \\
    \hat{\sigma}^{(0)}&=\frac{1}{N_s}\operatorname{tr}\left(\left(\widehat{\mathbf{M}}^{(0)}\right)^{-1}\left(\mathbf{z}\mathbf{z}^\dagger-\boldsymbol{S}\right)\right).
\end{align}
\end{subequations}
 
For the $t$-th iteration, with parameters 
$(\Theta^{(t)},\Phi^{(t)},\alpha^{(t)},\sigma^{(t)})$, we first fix the angles and update $\alpha$ and $\sigma$:
\begin{subequations}\label{eq:30}
\begin{align} \hat{\alpha}^{(t+1)}&=\frac{\mathbf{p}^\dagger\left(\mathbf{M}^{(t)}\right)^{-1}\mathbf{z}}{\mathbf{p}^\dagger(\mathbf{M}^{(t)})^{-1}\mathbf{p}}, \\
    \hat{\sigma}^{(t+1)}&=\frac{1}{N_s}\operatorname{tr}\left(\left(\mathbf{M}^{(t)}\right)^{-1}\left(\mathbf{z}\mathbf{z}^\dagger-\boldsymbol{S}\right)\right).
\end{align}
\end{subequations}
The covariance is then updated as $\widehat{\mathbf{M}}^{(t)}=\hat{\sigma}^{(t)}\sum_{i=1}^{N_{s}}\overline{\mathbf{v}}\left(\hat{\phi}_{i}^{(t)},\hat{\theta}_{i}^{(t)}\right)\overline{\mathbf{v}}\left(\hat{\phi}_{i}^{(t)},\hat{\theta}_{i}^{(t)}\right)^{\dagger}$. Next, with $\alpha$ and $\sigma$ fixed, the angle parameters are updated. Since the DOD and DOA are not paired as in the $H_2$ case, $\Theta$ and $\Phi$ must be updated separately. Taking $\Theta$ as an example, the coarse angle estimate vector is $\bar{\Theta}^{(t)}=[\theta_1^{(t)},\phi_1^{(t)},...,\theta_{N_s}^{(t)},\phi_{N_s}^{(t)}]^T$. The LM method is still employed for the iteration, but with the cost function now being $F(\bar{\Theta})=\|\bar{\mathbf{z}}-\alpha^{(t+1)}\mathbf{p}\|_{\mathbf{M}^{-1}}^{2}+\log\det(\mathbf{M})$, where $\|x\|_{A}^{2}=x^{\dagger}Ax$. The residual is then calculated as
\begin{equation}
e^{(t)}(\Theta)=\bar{\mathbf{z}}-\hat{\alpha}^{(t)}\mathbf{p}(\bar{\Theta}^{(t)}).
\label{eq:31}
\end{equation}
Similar steps yield $e^{(t)}(\Phi)$, and the residual difference is calculated as
\begin{equation}
e^{(t)}=\|e^{(t)}(\Theta)\|_2-\|e^{(t)}(\Phi)\|_2.
\label{eq:32}
\end{equation}
The iterative process terminates when the condition $\|e^{(t-1)}\|_2-\|e^{(t)}\|_2\leq\varepsilon_3$ is satisfied under the constraint $\|\boldsymbol{S}^{1/2}\widehat{\mathbf{M}}^{-1}\boldsymbol{S}^{1/2}-\boldsymbol{I}\|_2\leq\hat{\epsilon}_{\mathrm{EL}}$.

The stopping threshold $\varepsilon_1$ bounds the residual norm, whereas $\varepsilon_2$ and $\varepsilon_3$ bound successive decreases in their respective stopping quantities. In the experiments, they are selected on the validation set and then fixed for all test conditions. Similarly, $\delta$ and $\delta_{r}$ are determined by validation to balance the false split between specular and paired multipath components. Unless otherwise stated, all thresholds are kept unchanged across the synthetic, FDTD-simulated, and measured datasets to avoid test set tuning.

\subsection{The KLIC Adaptive Decision Framework}
\begin{figure}[!b]
\centerline{\includegraphics[width=22pc]{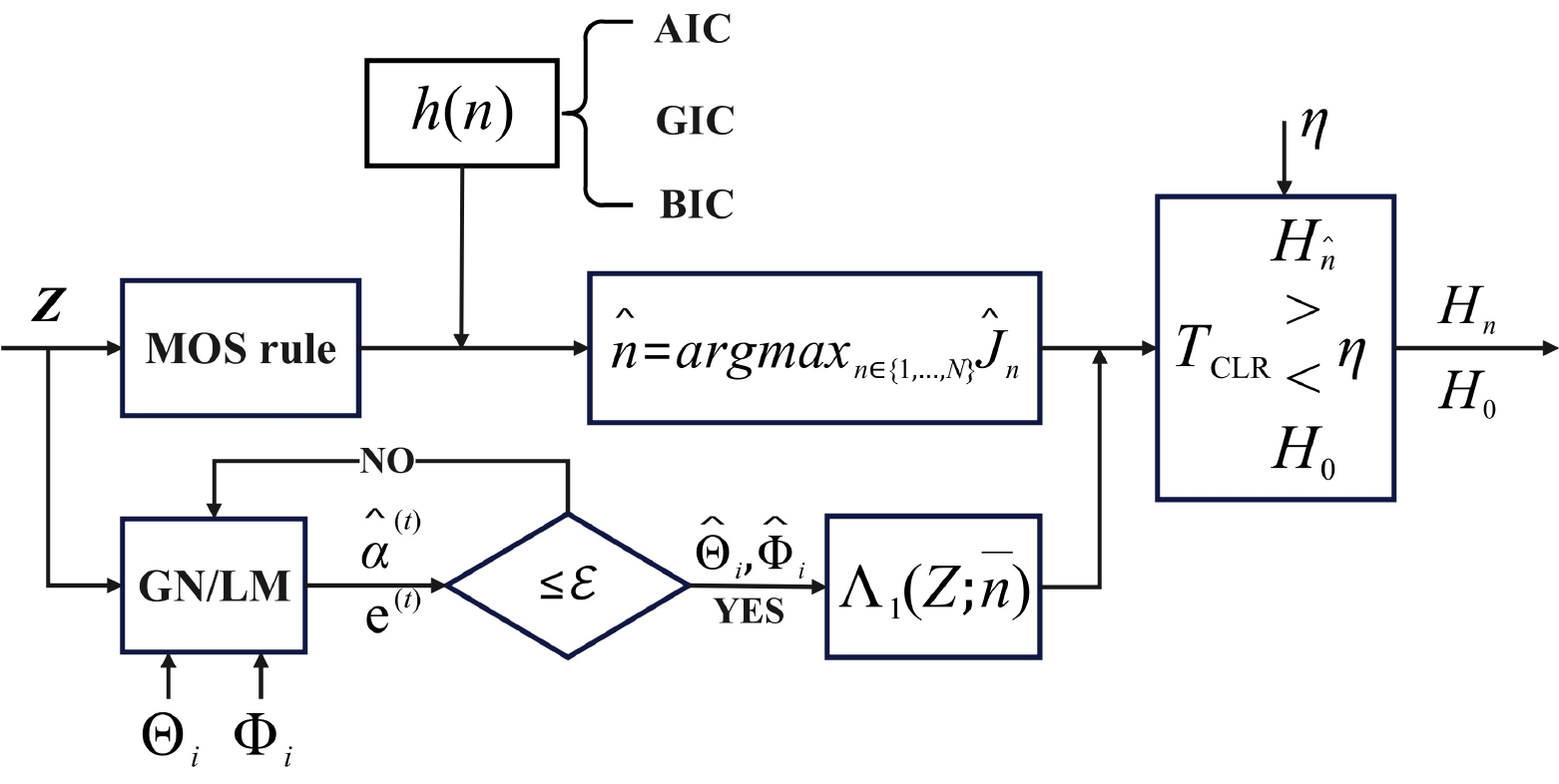}}
\caption{Block scheme of the adaptive detector.}
\label{fig3}
\end{figure}

In the general case, the number of paths $k_i$ and the angle parameters $\Theta_i, \Phi_i$ are all unknown. Following the same approach used to derive the MOS rule, we use a Taylor series expansion of the cross-entropy. By optimizing over $k_i$ to eliminate its dependency, this becomes equivalent to minimizing the KLD between the true PDF $\bar{f}$ and the model PDF $f_{n,i}$. Let $\hat{\Theta}_i$ and $\hat{\Phi}_i$ denote the estimates of $\Theta_i$ and $\Phi_i$ under these assumptions, respectively, thereby obtaining the decision rule in terms of the reasonable expectation of the unknown $\bar{f}$:
\begin{equation}
\begin{matrix}
E_{\bar{f}}\left [\log{{f_{n,i}}}\left (Z;\hat{\Theta}_i,\hat{\Phi}_i,k_i\right )\right ]-E_{\bar{f}}\left [\log{f_0}(Z;\hat{\Theta}_i,\hat{\Phi}_i)\right ] \begin{matrix} H_n \\ > \\ < \\ H_0 \end{matrix} 0 \\
\Rightarrow \quad\Lambda_{1}(Z;\bar{n}) = E_{\bar{f}}\left [\log{\frac{{f_{n,i}}\left (Z;\hat{\Theta}_i,\hat{\Phi}_i,k_i\right )}{{f_0}(\boldsymbol{Z};\hat{\Theta}_i,\hat{\Phi}_i)}}\right ] \begin{matrix} H_n \\ > \\ < \\ H_0 \end{matrix} 0.
\end{matrix}
\label{eq:33}
\end{equation}
The expectation with respect to the unknown PDF $\bar{f}$ is approximated using the fitted log-likelihood ratio
\begin{equation}
\Lambda_1(Z;\bar{n})=\log\frac{f_{n,i}\left(Z;\hat{\Theta}_i,\hat{\Phi}_i,k_i\right)}{f_0\left(Z;\hat{\Theta}_i,\hat{\Phi}_i\right)}.
\label{eq:34}
\end{equation}
Let $h(\bar{f})$ be the differential entropy of $\bar{f}$ and $h(\bar{f},f_{n,i})$ be the cross-entropy between $\bar{f}$ and $f_{n,i}$. We define $\mathcal{J}_n(\Theta_i,\Phi_i)=-h(\bar{f},f_{n,i})$ and denote its estimator by $\widehat{\mathcal{J}}_n$. Several feasible choices for $\widehat{\mathcal{J}}_n$ exist, namely:
\begin{subequations}\label{eq:35}
\begin{align} 
    \bullet \quad \widehat{\mathcal{J}}_n &= \log f_{n,i}\left(Z;\hat{\Theta_i},\hat{\Phi_i},k_i\right)-(k_i+k_s) \label{eq:35a} \\
    \bullet \quad \widehat{\mathcal{J}}_n &= \log f_{n,i}\left(Z;\hat{\Theta_i},\hat{\Phi_i},k_i\right)-\frac{1+\rho}{2}(k_i+k_s) \label{eq:35b} \\
    \bullet \quad \widehat{\mathcal{J}}_n &= \log f_{n,i}\left(Z;\hat{\Theta_i},\hat{\Phi_i},k_i\right)-\frac{k_i+k_s}{2}\log K+C 
\label{eq:35c}
\end{align}
\end{subequations}

Here, $K$ is the number of independent snapshots in $Z$, and $C$ is independent of the candidate model. For each $n$, $i$ denotes the configuration maximizing the corresponding penalized score. Finally, an estimate of $n$ is obtained as follows:
\begin{equation}
\hat{n}=\operatorname{arg\max}_{n\in\{1,...,N\}}\widehat{\mathcal{J}}_n.
\label{eq:36}
\end{equation}

This leads to the adaptive decision rule:
\begin{equation}
{{\widehat{\mathcal{J}}}_{\hat{n}}}-\log{f_0}(Z;\hat{\Theta_i},\hat{\Phi_i})\begin{matrix}{H_{\hat{n}}}\\>\\<\\H_0\end{matrix}0.
\label{eq:37}
\end{equation}

By introducing a threshold $\eta$ to control the probability of false alarm, the rule can be rewritten as

\begin{equation}
\begin{aligned}
&\max_{n\in\{1,...,N\}}\left \{{\log{{f_{n,i}}}\left ({Z;\hat{\Theta_i},\hat{\Phi_i},k_i}\right )-h(n)}\right \} \\
&\qquad  \qquad -\log{f_0}(Z;\hat{\Theta_i},\hat{\Phi_i})\begin{matrix}{H_{\hat{n}}}\\>\\<\\H_0\end{matrix}\eta \\
&\quad \Rightarrow\quad\max_{n\in\{1,...,N\}}\left \{{{\Lambda_1}(Z;\bar{n})-h(n)}\right \}\begin{matrix}{H_{\hat{n}}}\\>\\<\\H_0\end{matrix}\eta,
\label{eq:38}
\end{aligned}
\end{equation}
where $h(n)$ is a penalty term inherited from the MOS rule:
\begin{equation}
h(n)=
\begin{cases}
\left(k_i+k_s\right), & \text{AIC} \\
\frac{1+\rho}{2}\left(k_i+k_s\right),\ \rho>1, & \text{GIC} \\
\frac{k_i+k_s}{2}\log K, & \text{BIC}
\end{cases}
\label{eq:39}
\end{equation}

These penalties correspond to the Akaike Information Criterion (AIC), Generalized Information Criterion (GIC), and Bayesian Information Criterion (BIC), respectively. By adaptively adjusting the penalty term, we obtain the compressed log-likelihood ratio (CLR) detector. For sufficiently large $K$, this detector can be approximated by a suitably designed generalized likelihood ratio test (GLRT). The resulting decision rule provides the foundation for a detection architecture capable of discriminating among one null hypothesis and multiple alternatives, as illustrated in Fig. \ref{fig3}.

\subsection{DEKL Method}
Building on the parameter estimation and decision testing pipeline developed in the previous section, we design a classification and decision neural network, KLNet, for MOS in subsampling systems. Grounded in the KLIC-based decision framework, KLNet follows the decision logic illustrated in Fig. \ref{fig2} by mapping the covariance-domain echo representation to feature extraction, penalty-aware weighting, model-order probability estimation, and confidence-based hypothesis selection. We term this KLIC-guided classification framework DEKL. Its KLNet inference directly predicts the hypothesis without rerunning the parameter-estimation loops in Section III-A.

Although the KLIC framework provides a theoretically grounded decision rule, its practical deployment faces several limitations: (i) reliance on handcrafted statistical features and assumed parametric distributions (e.g., complex Gaussian), rendering it sensitive to model mismatch and non-Gaussian interference; (ii) exhaustive grid search over angle parameters and iterative likelihood evaluation incur prohibitive computational costs; and (iii) fixed penalty $h(n)$ and decision threshold $\eta$ limit adaptability to varying channel conditions. DEKL addresses these limitations by using KLNet to learn covariance-domain representations and adaptive decision scores from data, while the KLIC rule in \eqref{eq:38} guides the design of the penalty-aware attention and confidence-based decision modules.

The covariance-domain input follows from the spatial structure of the models in Section II. For fixed path geometry, \eqref{eq:6} and \eqref{eq:9} have the form $\mathbf z=\mathbf G\mathbf b+\mathbf r$, where $\mathbf G$ collects the effective array responses and $\mathbf b$ contains the path amplitudes. With zero-mean noise uncorrelated with $\mathbf b$, expanding the outer product and taking expectations gives
\begin{equation}
\begin{aligned}
\mathbf R_z&=\mathbb E[\mathbf z\mathbf z^\dagger]
=\mathbf G\mathbf Q\mathbf G^\dagger+\mathbf R_r\\
&=\sum_{j=1}^{J}\sum_{m=1}^{J}Q_{jm}\mathbf g_j\mathbf g_m^\dagger+\mathbf R_r,
\end{aligned}
\label{eq:40}
\end{equation}
where $\mathbf Q=\mathbb E[\mathbf b\mathbf b^\dagger]$, $\mathbf R_r=\mathbb E[\mathbf r\mathbf r^\dagger]$, and $\mathbf g_j$ is the $j$-th of the $J$ columns of $\mathbf G$. For deterministic amplitudes, $\mathbf Q=\mathbf b\mathbf b^\dagger$. Thus, the uncentered second moment $\mathbf R_z$, estimated by $\boldsymbol S$ below, includes coherent mean components as well as random fluctuations. Under $H_1$, the array responses obey the equal-angle constraint. Under $H_2$, the paired paths introduce additional angular responses and correlation terms. Under $H_3$, the diffuse covariance in \eqref{eq:12}--\eqref{eq:13} aggregates contributions from distributed scattering directions. These structures provide physical cues for distinguishing the candidate models through spatial second-order statistics.

For the proposed KLNet input interface, let $\boldsymbol{S}=K^{-1}\sum_{k=1}^{K}\mathbf{z}_k\mathbf{z}_k^\dagger\in\mathbb{C}^{M_TM_R\times M_TM_R}$ denote the full transmit--receive covariance. The receive-domain descriptor is defined as
\begin{equation}
\boldsymbol{R}_{\mathrm R}=\frac{1}{M_T}\sum_{t=1}^{M_T}\boldsymbol{S}_{tt}\in\mathbb{C}^{M_R\times M_R},
\label{eq:41}
\end{equation}
where $\boldsymbol{S}_{tt}$ is the receive-channel diagonal block associated with transmitter $t$. For receive-channel observations, $\boldsymbol{R}_{\mathrm R}$ is estimated directly. A fixed direction-cosine grid $u_q=-1+q/8$, $q=0,\ldots,15$, defines the receive steering matrix $\boldsymbol{A}_{\mathrm R}\in\mathbb{C}^{M_R\times16}$, with entries $[\boldsymbol{A}_{\mathrm R}]_{mq}=M_R^{-1/2}e^{j2\pi d_m u_q/\lambda}$ for receive-element position $d_m$. The beam-output covariance and network input are
\begin{equation}
\begin{aligned}
\boldsymbol{B}&=\boldsymbol{A}_{\mathrm R}^{\dagger}\boldsymbol{R}_{\mathrm R}\boldsymbol{A}_{\mathrm R}\in\mathbb{C}^{16\times16},\\
\mathcal{X}&=\operatorname{stack}\bigl(\Re\{\boldsymbol{B}\},\Im\{\boldsymbol{B}\}\bigr)\in\mathbb{R}^{2\times16\times16}.
\end{aligned}
\label{eq:42}
\end{equation}
The diagonal entries of $\boldsymbol B$ represent receive-beam powers, while its off-diagonal entries describe correlations between beams. Both axes of $\boldsymbol{B}$ index receive directions. The grid standardizes the input dimensions while retaining the resolution limits of each receive aperture. Table~\ref{tab_input_dimensions} summarizes the dimensions. Each sample is assigned a one-hot hypothesis label for supervised classification.

\begin{table}[!t]
\caption{Dimensions of the Proposed KLNet Input Interface}
\label{tab_input_dimensions}
\centering
\begin{tabular}{lcc}
\hline
Representation & Simulation & Measurement \\
\hline
Tx $\times$ Rx & $10\times10$ & $2\times4$ \\
Full model $\boldsymbol{S}$ & $100\times100$ & $8\times8$ \\
Receive descriptor $\boldsymbol{R}_{\mathrm R}$ & $10\times10$ & $4\times4$ \\
Network input $\mathcal{X}$ & $2\times16\times16$ & $2\times16\times16$ \\
\hline
\end{tabular}
\end{table}

The encoder comprises three sequential blocks, each containing a convolutional layer, Batch Normalization (BN), and ReLU activation, followed by max pooling and global average pooling. Shallow layers capture local correlation patterns, such as localized diagonal dominance associated with specular reflection, while deeper layers aggregate global signatures, including angular offset clusters characteristic of multipath. Here, these patterns describe energy concentration and angular separation in the receive-beamspace covariance. The attention module emulates the MOS penalty mechanism by computing channel-wise weights that emphasize features associated with the most probable hypothesis, thereby approximating the adaptive penalty selection in \eqref{eq:39}. The network learns these weights without explicitly computing $h(n)$.

The decoder adopts a recurrent fully connected block with $N_h$ hidden layers, each containing $N_u$ neurons with ReLU activations. KLNet employs three $3\times3$ convolutional blocks with 8, 16, and 32 channels, a squeeze-and-excitation MOS attention module with a reduction ratio of 8, and two fully connected decoder layers with 24 neurons. With shared iterative decoder weights, the complete network contains 7,881 trainable parameters, including 6,072 in the convolution--BN encoder, 292 in the attention module, 1,392 in the decoder, and 125 in the classification and confidence heads. This iterative refinement (2--3 iterations until convergence) mirrors the parameter estimation loops in Section III-A. The output bifurcates into two branches: (i) a classification branch with a softmax layer producing $z(\ell)$ for $\ell=0, \ldots, L_{\max}$, trained by maximizing the log-likelihood $\underset{w}{\operatorname{max}} \ln(z(\ell^*|x;w))$; and (ii) a confidence branch with a sigmoid mapping yielding a confidence score in $[0, 1]$ used to gate further refinement. The two branch outputs are passed to the decision module, which outputs the final hypothesis. Through this architecture, DEKL unifies the interpretability of KLIC with the flexibility of deep learning for robust multipath classification.

\begin{equation}
\mathcal L_{\rm cls}(w)=-\frac{1}{N_{\rm tr}}\sum_{i=1}^{N_{\rm tr}}
\log z(\ell_i^*\mid x_i;w).
\label{eq:43}
\end{equation}
Here, $N_{\rm tr}$ is the number of training samples, $x_i$ denotes the covariance-domain network input, and $\ell_i^*\in\{0,1,2,3\}$ is its hypothesis label. The sigmoid output in Fig.~\ref{fig2} can be written as
\begin{equation}
c_i=\operatorname{sigmoid}\!\left(e_i^{(T)}\right)
=\frac{1}{1+\exp\!\left(-e_i^{(T)}\right)},
\label{eq:44}
\end{equation}
where $e_i^{(T)}$ is the input to the confidence-branch sigmoid at decoder iteration $T$. The classification probabilities and confidence score are the two inputs to the decision module.

For the proposed joint training, all samples undergo three decoder iterations, with $\mathcal L_{\rm cls}$ evaluated at the third output. Let $\widehat\ell_i^{(t)}=\arg\max_\ell z_i^{(t)}(\ell)$, with class-score ties resolved by the smallest index. The confidence target $r_i^{(t)}=\mathbb I\{\widehat\ell_i^{(t)}=\ell_i^*\}$ is treated as constant during backpropagation. The existing confidence head is supervised at iterations two and three by
\begin{equation}
\begin{aligned}
\mathcal L_{\rm conf}=-\frac{1}{2N_{\rm tr}}
\sum_{i=1}^{N_{\rm tr}}\sum_{t=2}^{3}
\bigl[&r_i^{(t)}\log c_i^{(t)}\\
&+(1-r_i^{(t)})\log(1-c_i^{(t)})\bigr],
\end{aligned}
\label{eq:45}
\end{equation}
where $z_i^{(t)}(\ell)$ and $c_i^{(t)}$ are the branch outputs at iteration $t$. The joint objective is $\mathcal L=\mathcal L_{\rm cls}+\mathcal L_{\rm conf}$.

\begin{algorithm}[t]
\normalsize
\caption{Confidence-Gated DEKL Inference}
\label{alg:dekl_inference}
\begin{algorithmic}[1]
\REQUIRE Covariance-domain input $\mathcal X_i$; trained KLNet; validation-selected threshold $\tau_c$.
\ENSURE Final hypothesis $\widehat H_i$.
\STATE Evaluate the decoder and both heads at $t=1,2$.
\IF{$\widehat\ell_i^{(2)}=\widehat\ell_i^{(1)}$ and $c_i^{(2)}>\tau_c$}
\STATE $\widehat H_i \leftarrow H_{\widehat\ell_i^{(2)}}$.
\ELSE
\STATE Perform the third refinement.
\STATE $\widehat H_i \leftarrow H_{\widehat\ell_i^{(3)}}$.
\ENDIF
\RETURN $\widehat H_i$.
\end{algorithmic}
\end{algorithm}

At inference, the classification branch supplies the candidate hypothesis and the confidence branch gates further refinement:
\begin{equation}
\begin{aligned}
T_i&=\begin{cases}
2,&\widehat\ell_i^{(2)}=\widehat\ell_i^{(1)}\ \text{and}\ c_i^{(2)}>\tau_c,\\
3,&\text{otherwise},
\end{cases}\\
\widehat H_i&=H_{\widehat\ell_i^{(T_i)}}.
\end{aligned}
\label{eq:46}
\end{equation}
Convergence here denotes a stable hypothesis with sufficient confidence. At the three-iteration limit, the current class is returned regardless of confidence, retaining every sample in the four-class evaluation.

On a validation set $\mathcal D_v$ disjoint from training and testing, $\tau_c$ maximizes the accuracy of \eqref{eq:46} over $\{0,1\}\cup\{c_i^{(2)}:i\in\mathcal D_v\}$. Ties favor fewer mean iterations, then the larger threshold. The strict comparison makes $\tau_c=1$ the three-iteration reference. The selected threshold is fixed before testing; it implements confidence gating without identifying its numerical value with the statistical threshold $\eta$ in \eqref{eq:38}.

\section{SIMULATION AND EXPERIMENTS}

\begin{figure}[!t]
\centerline{\includegraphics[width=0.43\textwidth]{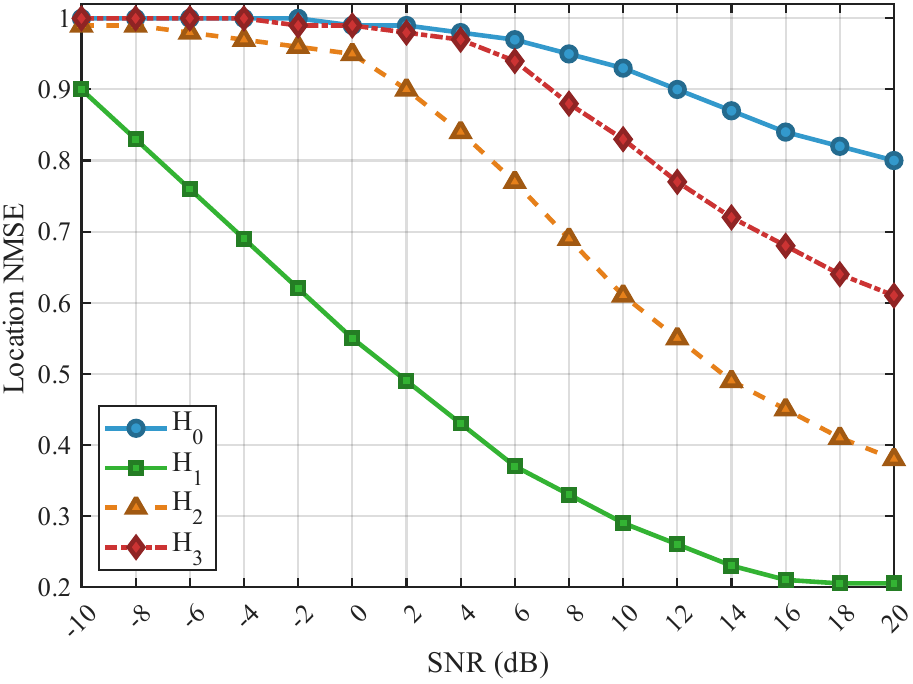}}
\caption{Positioning NMSE under the error model.}
\label{fig4}
\end{figure}

In low-SNR and low-snapshot regimes, DOA estimation is often affected by outliers, leading to a significant increase in the positioning normalized mean-square error (NMSE), as shown in Fig. \ref{fig4}. In the pure-noise case, spurious targets may be assigned random spatial positions, yielding the largest errors. Imposing the specular-reflection angle constraint effectively reduces the angular parameter dimension and suppresses estimation variance, whereas diffuse scattering introduces multiple random paths, spreads the echo energy over spatial frequencies, and increases model ambiguity. These observations motivate adaptive model-order selection and scattering-type classification before subsequent target parameter recovery. Accordingly, the DEKL training process adjusts the network weights $w$ to maximize the posterior probability of the correct model order $\ell^*$ for a given covariance-domain input $x$, which can be regarded as a data-driven approximation to the KL-GLRT estimator under model mismatch and limited snapshots.

\begin{figure}[!b]
\centerline{\includegraphics[width=21pc]{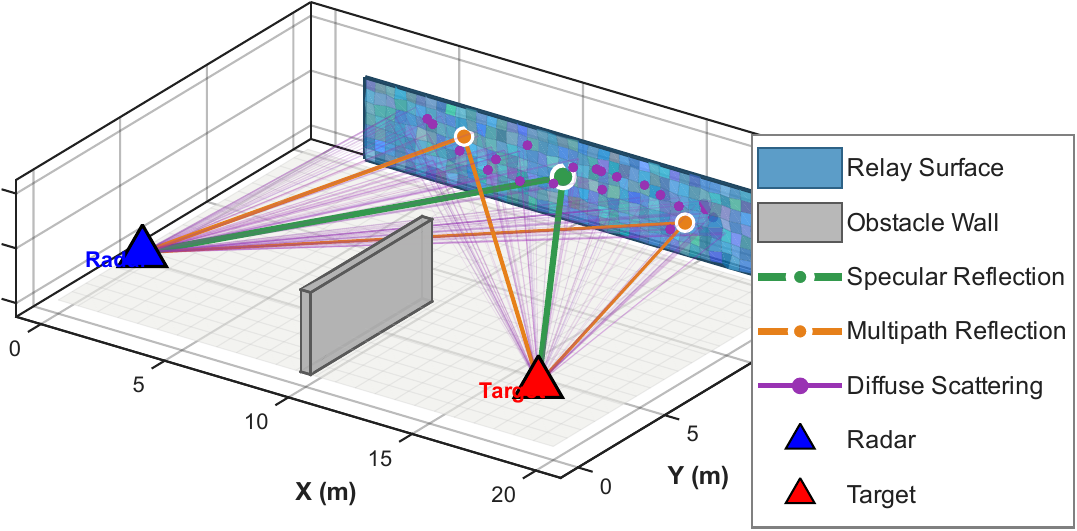}}
\caption{High-fidelity electromagnetic simulation environment for independent test set generation.}
\label{fig5}
\end{figure}

\begin{figure}[!t]
\centerline{\includegraphics[width=21pc]{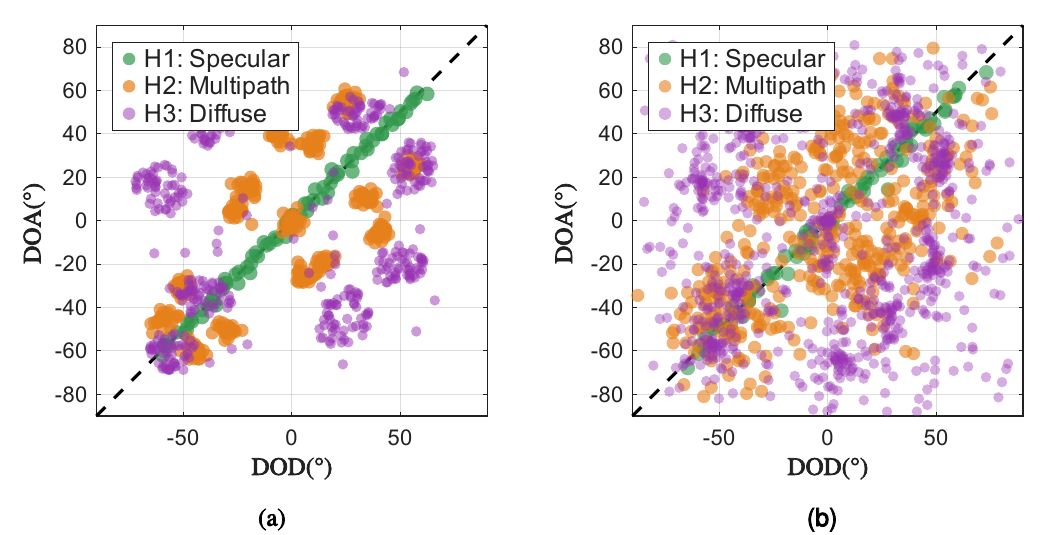}}
\caption{Comparison of DOD--DOA feature space distributions: \\
(a) training set; (b) test set.}
\label{fig6}
\end{figure}

\begin{figure*}[!b]
\centering
\subfloat[]{\includegraphics[width=0.45\textwidth]{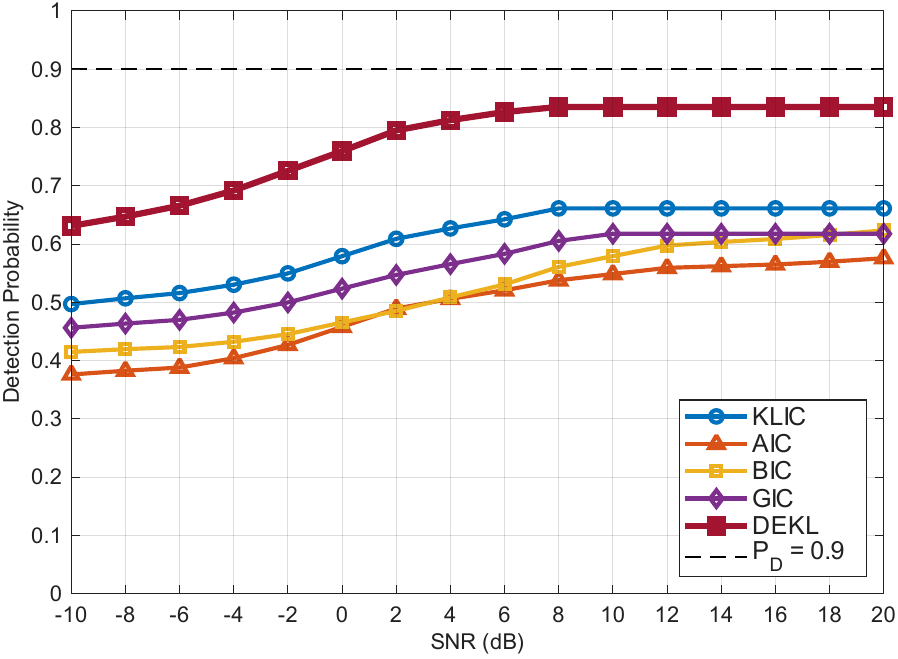}
\label{PDSNR}}
\hfil
\subfloat[]{\includegraphics[width=0.45\textwidth]{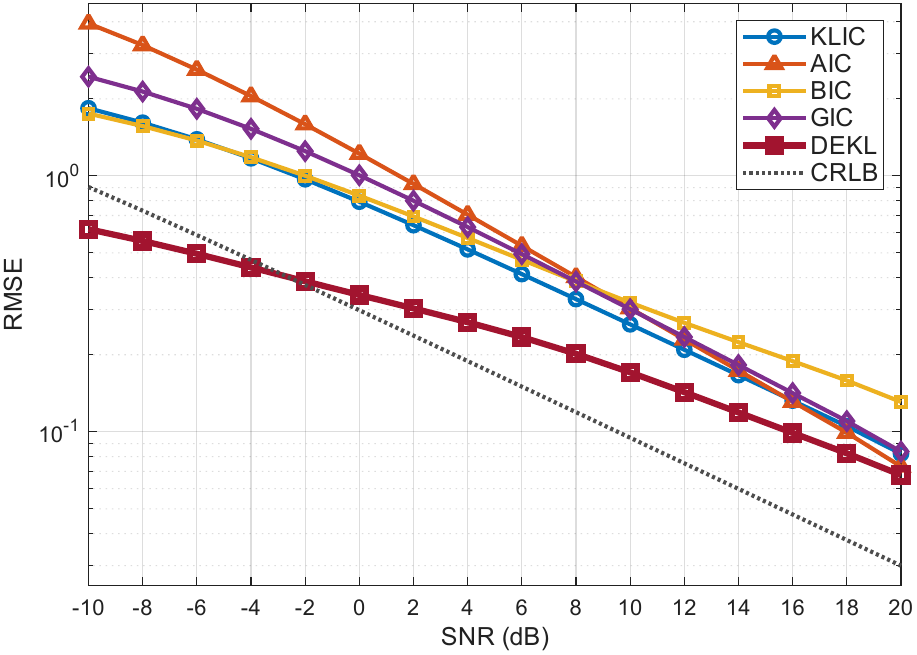}
\label{RMSESNR}}
\caption{Performance comparison at a fixed false alarm rate. (a) Detection probability (\(P_D\)). (b) Root-mean-square error (RMSE).}
\label{fig7}
\end{figure*}

\subsection{Simulation Settings}

Based on the physics-based modeling framework established in Section II, a synthetic dataset encompassing four multipath scenarios is generated using MATLAB, mainly serving as the training set for the proposed network. The radar operates at a frequency of $f_c = 77$ GHz, with $M_T = 10$ transmit antennas, $M_R = 10$ receive antennas, and $L = 10$ pulses. The noise variance is normalized to $\sigma^2=1$. Four hypotheses are considered: pure noise ($H_0$), specular reflection ($H_1$), multipath reflection ($H_2$), and diffuse scattering ($H_3$). For each hypothesis, 2,000 samples are generated for training and validation. To avoid category bias, each training batch is constructed with a uniform distribution of samples across different model orders.

To rigorously address the ``inverse crime'' issue and assess generalization capability in nonideal physical environments, a high-fidelity electromagnetic simulation scenario is constructed using the electromagnetic simulation software gprMax, as depicted in Fig. \ref{fig5}, with the system configuration consistent with that in Fig. \ref{fig1}. This simulation explicitly models the reflection and diffuse scattering effects of different relay surfaces through material and surface roughness parameters. The finite-difference time-domain (FDTD) method in gprMax is employed to calculate electromagnetic wave propagation, scattering, diffraction, and multipath interactions, thereby yielding an independent test set with high physical realism for algorithm evaluation under distribution shift.

Fig. \ref{fig6} illustrates significant divergence in physical feature distribution between the synthetic training set and the FDTD-simulated test set. Analytically modeled training data (Fig. \ref{fig6}(a)) exhibit ideal statistical regularity, with data points tightly clustered around the theoretically predicted geometric manifold. In contrast, the FDTD-simulated test data (Fig. \ref{fig6}(b)) displays substantial dispersion and randomness, characterized by numerous outliers and irregular clusters resulting from incoherent scattering induced by surface microstructures.

\subsection{Simulation Results}

To evaluate the performance gain of the proposed decision scheme, we benchmark the algorithm against several baselines, including traditional information-criterion-based decision rules (AIC, BIC, GIC) and the previous standard KLIC-based adaptive decision algorithm. The synthetic dataset trains DEKL and determines validation parameters, while the FDTD-simulated dataset serves as an independent test set to evaluate generalization under physical mismatch.

In Fig. \ref{fig7}(a), a reference line is set at $P_D = 0.9$, representing successful classification in engineering terms. As SNR increases, detection probability improves for all methods. DEKL achieves high detection even at low SNR, effectively exploiting discriminative features, whereas conventional methods show distinct inflection points, with KLIC appearing earlier due to its adaptive advantage. In the medium-to-high SNR range, all methods approach their performance limits, but DEKL maintains a clear lead and stays closer to the reference threshold.

As shown in Fig. \ref{fig7}(b), the DEKL curve descends smoothly in the low-to-medium SNR range, mitigating threshold effects in parameter estimation and allowing errors to overcome noise dominance earlier, improving asymptotic convergence and approaching the CRLB faster. KLIC performance converges to DEKL at higher SNRs, yet a gap remains in the intermediate range. Other conventional methods, affected by model penalties or adaptation delays, show delayed error reduction at low SNRs. Together, the two subplots illustrate that in the low-to-medium SNR range, DEKL simultaneously increases detection probability and reduces parameter error, enhancing real-time adaptive signal reconstruction and multiclass signal discrimination. KLIC demonstrates second-best performance, whereas other methods exhibit detection--estimation asynchrony, adversely affecting downstream tasks like tracking, positioning, imaging, or data fusion.

\begin{figure*}[!t]
\centering
\subfloat[]{\includegraphics[width=0.45\textwidth]{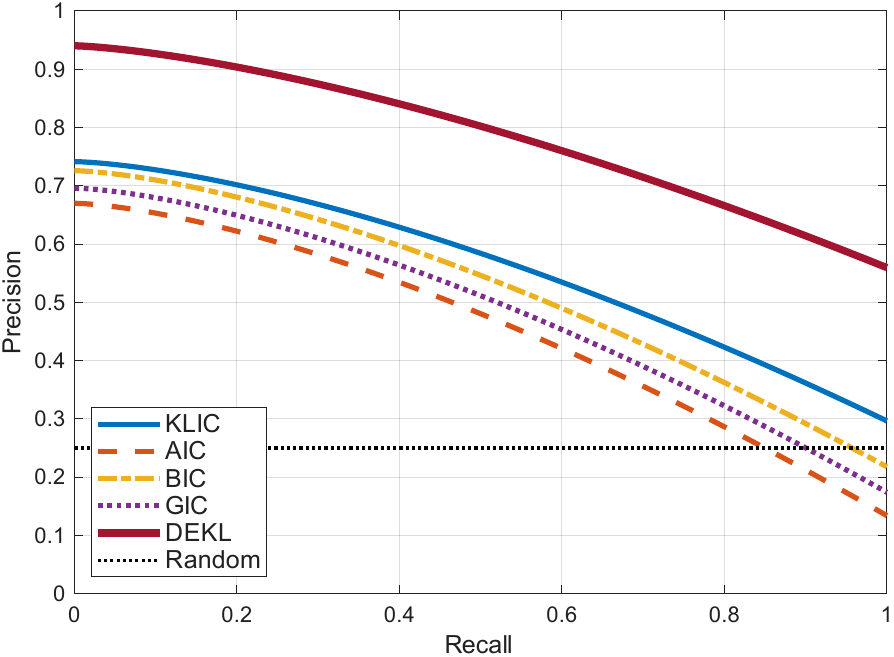}
\label{PR}}
\hfil
\subfloat[]{\includegraphics[width=0.45\textwidth]{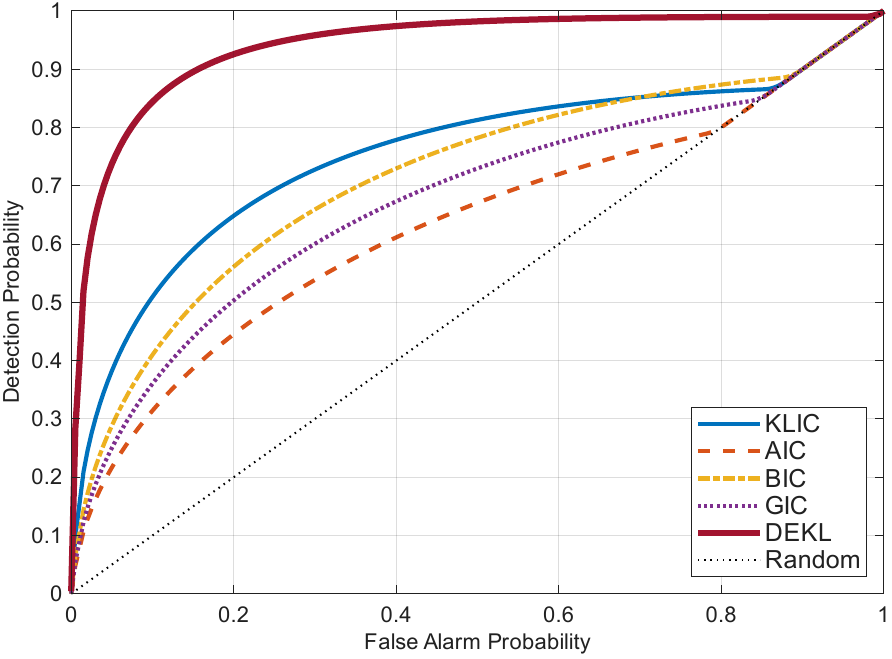}
\label{ROC}}
\caption{Performance evaluation using precision--recall (PR) and  receiver operating characteristic (ROC) Curves. (a) PR curve illustrates the tradeoff between precision and recall across different thresholds. (b) ROC curve visualizes the overall discriminative ability by plotting the true positive rate against the false positive rate.}
\label{fig8}
\end{figure*}

To evaluate overall multiclass performance, the one-vs-rest (OvR) strategy treats each class as positive while all others are negative. In Fig. \ref{fig8}(a), the horizontal ``Random'' baseline shows expected precision for a nondiscriminative classifier equal to class prevalence. In Fig. \ref{fig8}(b), the diagonal ``Random'' line represents an uninformative random classifier. In both, the DEKL curve forms the outer envelope, achieving high recall with minimal precision loss and low false alarm rate, indicating synergy between likelihood ranking and probability calibration for stronger discrimination and robust threshold response. Among information-theoretic methods, KLIC is the most robust baseline, maintaining optimal likelihood discrimination and stable performance across ranges, though a gap persists relative to DEKL. Other methods, affected by noise or penalties, show premature degradation or downward-shifted envelopes, reflecting poor ranking consistency.

\subsection{Experimental Evaluation}

\begin{figure*}[!t]
\centerline{\includegraphics[width=42pc]{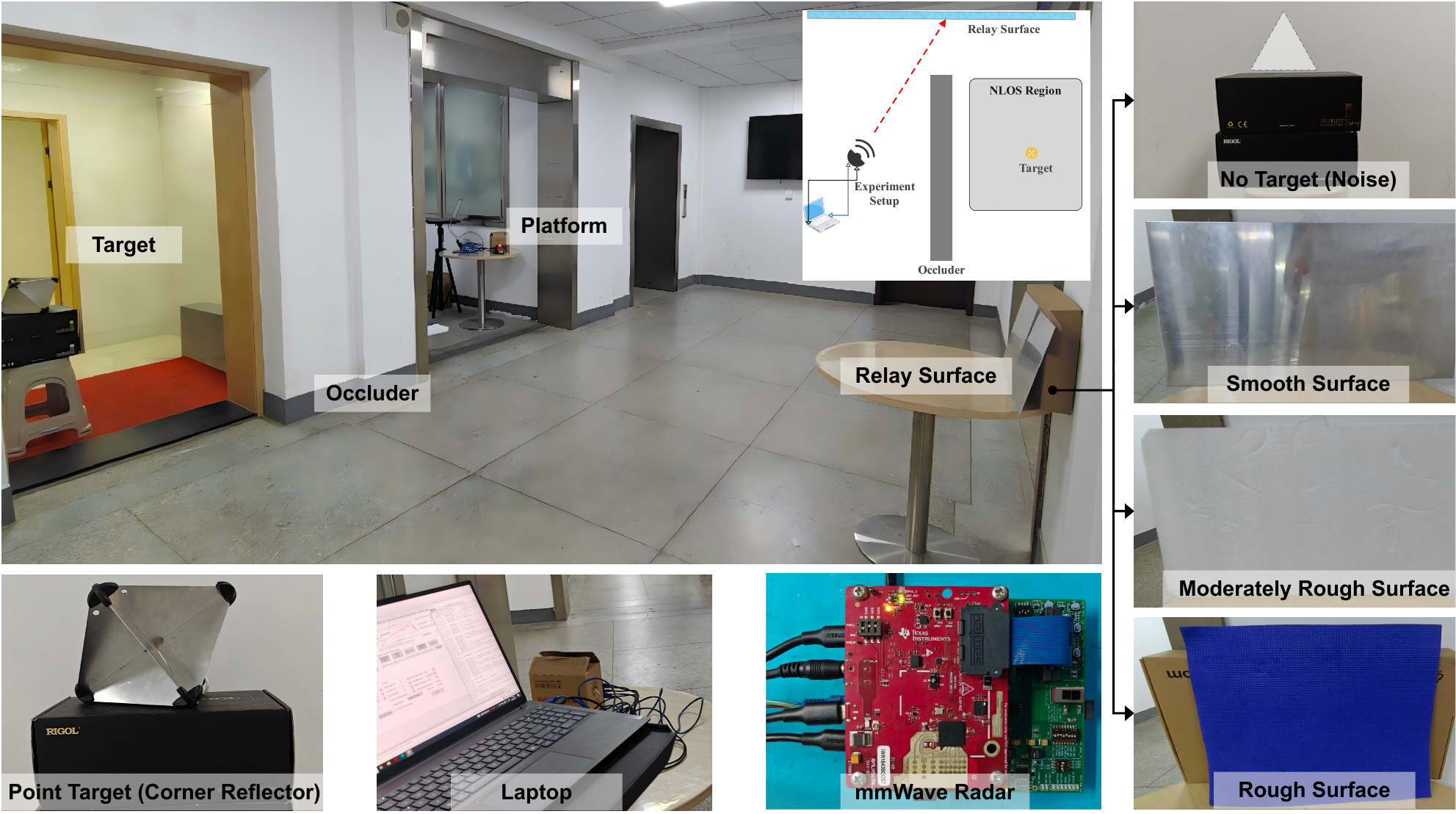}}
\caption{Measured setup for NLOS relay-surface verification includes a radar capture platform, an impenetrable nonreflective obstacle, a hidden target region, a corner reflector target, and representative relay surfaces for various scattering cases: no target, smooth, moderately rough, and rough.}
\label{fig9}
\end{figure*}

To further verify the practical applicability of the proposed framework, real-world measured radar echoes are used for independent evaluation \footnote{The term ``real-world measurements'' in this work refers to measurements acquired with radar hardware indoors. The corner reflector serves as a repeatable high-radar-cross-section surrogate for a vehicle target, while the stationary platform isolates relay-surface scattering and retains hardware nonidealities, calibration errors, and background clutter. Therefore, the experiment validates the proposed scattering classification stage; vehicle-mounted radar and cooperative perception systems are beyond the scope of this study.}. Unlike synthetic and FDTD-simulated data, measured echoes include hardware nonidealities, background clutter, system calibration errors, and nonideal effects induced by the relay-surface microstructure, thereby providing a more realistic test of algorithm robustness. Fig. \ref{fig9} shows the relay-assisted NLOS experimental scenario. The top-view layout follows Fig. \ref{fig1}: the radar platform on the left is aligned with the relay surface, a corner reflector is used as the target in the NLOS region on the right, and the direct path is blocked by the occluder, so useful echoes are generated entirely by relay-surface scattering. The right column shows the relay surfaces used in the experiment, corresponding to the $H_0$--$H_3$ cases, and also marks the point target, radar device, and host computer for data acquisition to ensure reproducibility.

\begin{table}[!t]
    \centering
    \caption{Measured experiment configuration}
    \begin{tabular}{c|c|c|c}
        \hline \hline
        Parameter & Value & Parameter & Value \\
        \hline
        Start frequency & 77 GHz & ADC samples & 256 \\
        Frequency slope & 30 MHz/$\mu$s & Chirps per frame & 8 \\
        Sampling rate & 6000 ksps & Frames per capture & 128 \\
        Ramp end time & 60 $\mu$s & Idle time & 10 $\mu$s \\
        Frame periodicity & 8 ms & Sampled BW & 1.28 GHz \\
        Number of Rx & 4 & Number of Tx & 2 \\
        \hline
    \end{tabular}
    \label{tab_measured_config}
\end{table}

The experiment uses a COTS millimeter-wave radar platform, including the TI IWR1843BOOST and DCA1000EVM. The radar has two transmit antennas with a horizontal spacing of two wavelengths and four receive antennas with a horizontal spacing of half a wavelength. For the measured dataset, 50 raw ADC capture files are collected for each case, yielding 200 classification samples in total. Each sample corresponds to one complete capture file, with its frames processed jointly rather than counted as separate classification examples. The binary data contain signed 16-bit I/Q values and are decoded into a $128\times8\times256\times4$ array ordered by frame, chirp, fast-time sample, and receive channel. A channel-wise complex mean is removed over each record before extracting range FFT waveforms, range-Doppler maps, and receive-channel covariance features. All comparison algorithms use the same feature-extraction pipeline to ensure fair evaluation. Table~\ref{tab_measured_config} summarizes the measured parameter configuration. The sampled bandwidth is $B_{\mathrm{samp}}=\mu N_{\mathrm{ADC}}/f_s=1.28$ GHz, where $\mu=30$ MHz/$\mu$s is the chirp slope and $N_{\mathrm{ADC}}/f_s=256/(6\,\mathrm{MHz})\approx42.67\,\mu$s is the ADC sampling window within the 60-$\mu$s ramp. Reference labels follow the recorded acquisition conditions shown in Fig. \ref{fig9}. The evaluation pertains to these relay-surface and target configurations.

\begin{figure*}[!t]
\centerline{\includegraphics[width=0.80\textwidth]{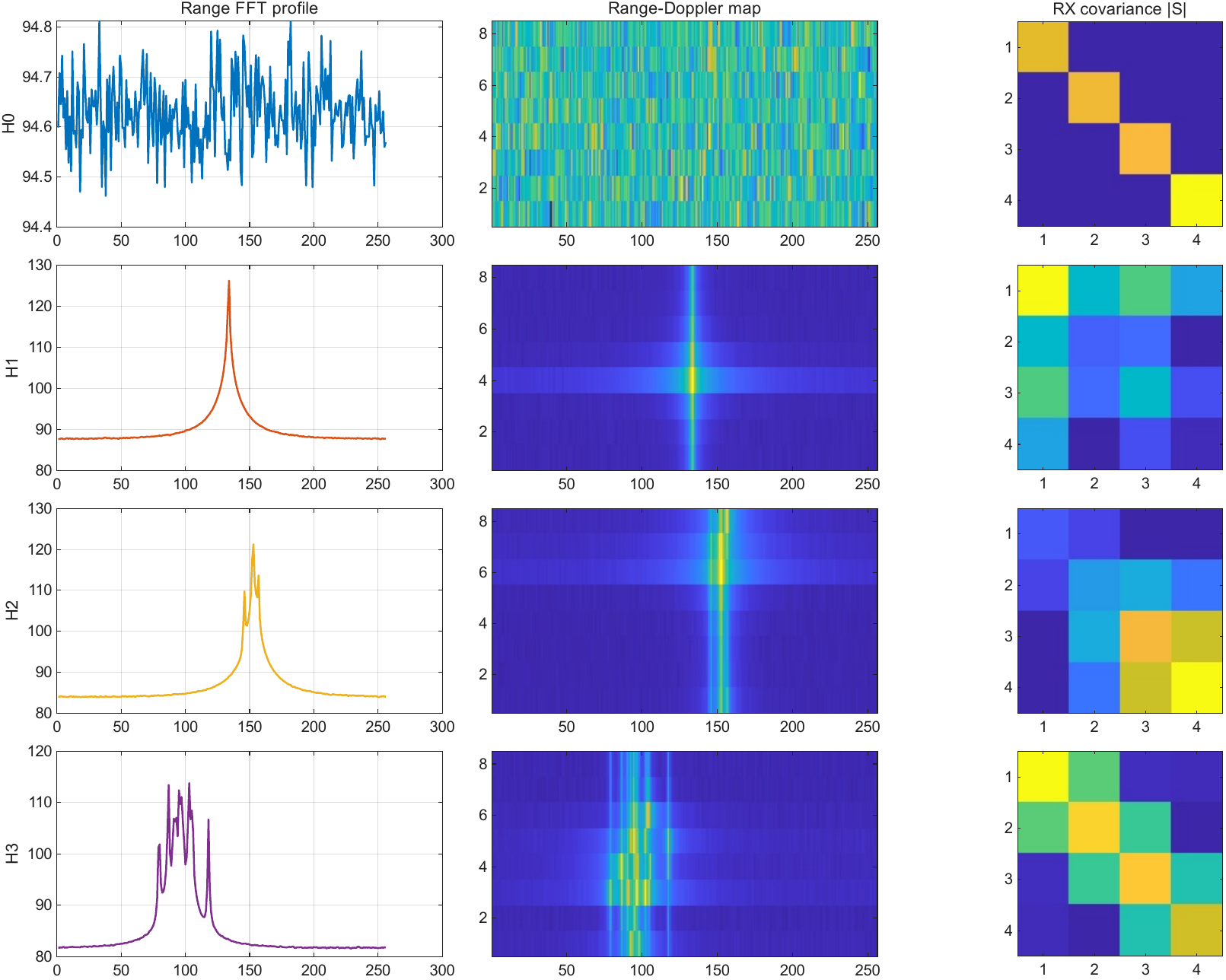}}
\caption{Representative multidomain input features under different relay-surface scattering hypotheses. Each row corresponds to a hypothesis from $H_0$ to $H_3$, and each column displays the range FFT profile, range-Doppler map, and spatial receive covariance matrix, respectively.}
\label{fig10}
\end{figure*}

\begin{figure}[!t]
\centerline{\includegraphics[width=0.43\textwidth]{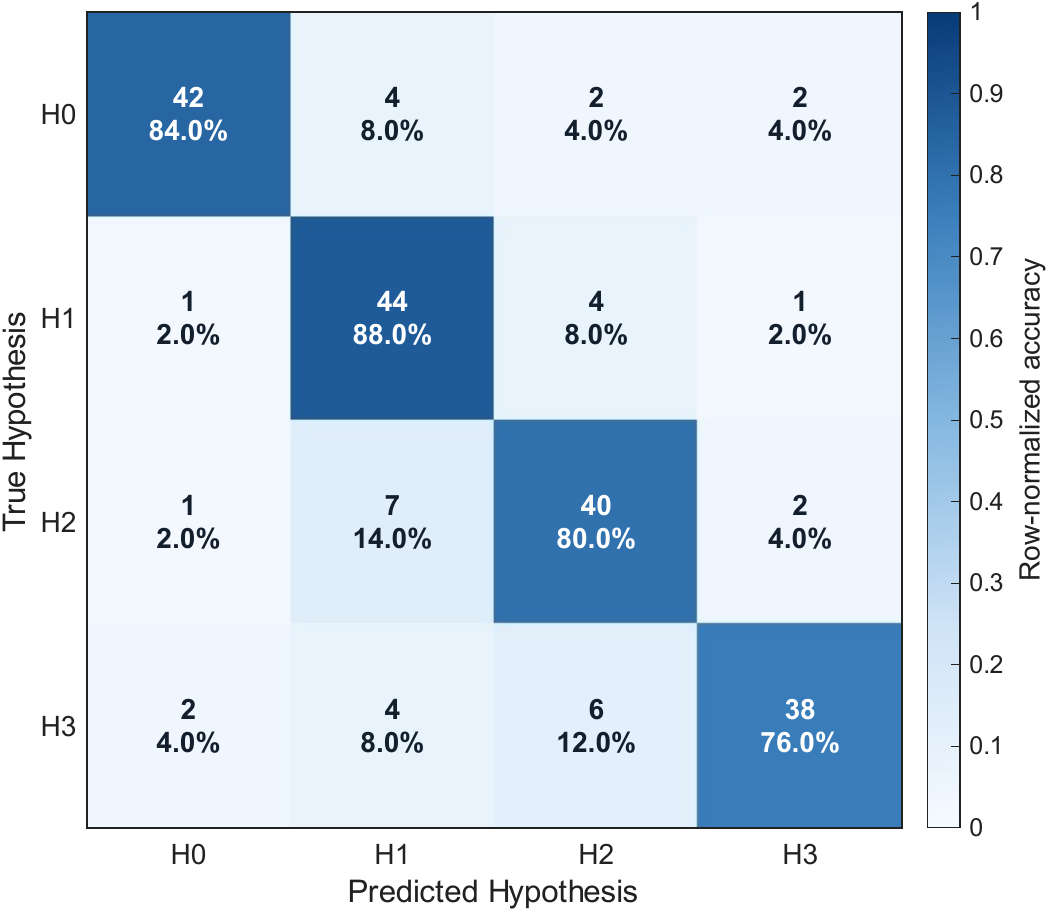}}
\caption{Confusion matrix of the proposed DEKL method on the measured test dataset.}
\label{fig11}
\end{figure}

\begin{figure}[!b]
\centerline{\includegraphics[width=0.43\textwidth]{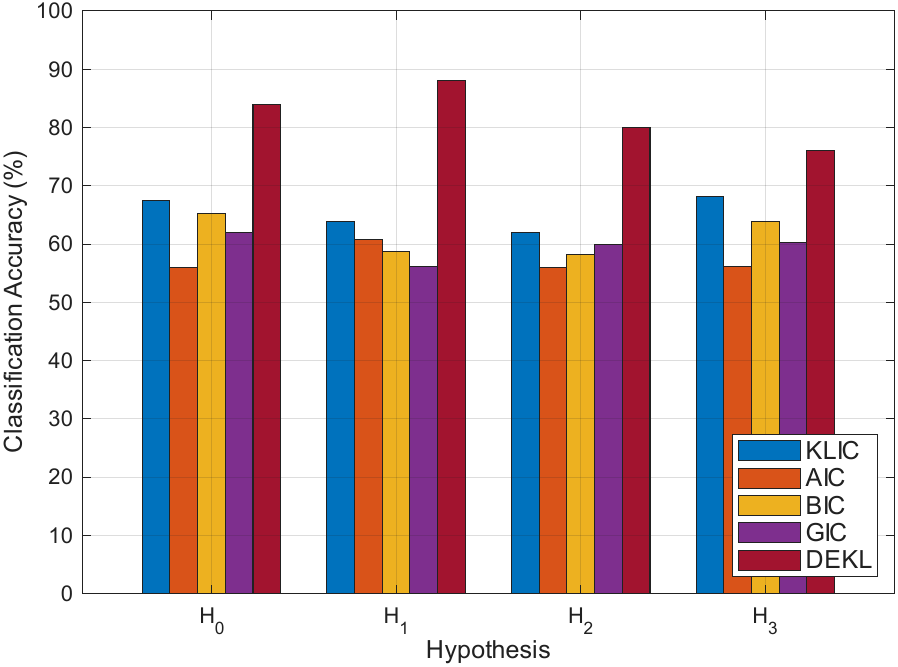}}
\caption{Per-hypothesis classification accuracy of the compared methods.}
\label{fig12}
\end{figure}

Fig. \ref{fig10} presents representative measured features for the four hypotheses, extracted from raw ADC samples. For $H_0$, the range FFT waveform fluctuates near the noise floor lacking a stable peak, and the range-Doppler map exhibits random energy. The covariance matrix contains predominantly weak diagonal components, consistent with noise-dominated echoes under the null hypothesis.
Under specular reflection ($H_1$), the range FFT waveform displays a clear, concentrated peak, and the range-Doppler map presents localized high-energy responses. The covariance matrix exhibits stronger spatial coherence, reflecting deterministic specular reflection signals, consistent with the $H_1$ model's stable main path and geometrically constrained angles.
For moderately rough surfaces ($H_2$), adjacent sidelobe peaks surround the main range FFT peak, and target range-cell energy broadens in the range-Doppler map. Unlike $H_1$, enhanced off-diagonal covariance components and uneven spatial correlations are consistent with multiple relay-surface reflections. These features match the $H_2$ model, where paired multipath components induce angular mismatch and added spatial correlations.
For highly rough diffuse scattering ($H_3$), the range FFT waveform is irregular and multipeaked, with range-Doppler energy spreading widely. The covariance matrix displays scattered correlations, reflecting strong randomness and broad spatial distributions. Unlike earlier hypotheses, $H_3$ features exhibit lower energy concentration and weaker determinism, matching the diffuse scattering assumption.

These maps illustrate differences among the echo structures under the four conditions. These observations support the use of covariance-domain features for discrimination in the collected dataset, without establishing a unique mapping between echo structure and surface roughness.

Fig. \ref{fig11} evaluates DEKL on measured data. For $H_0$ to $H_3$, DEKL correctly classifies 42, 44, 40, and 38 samples, respectively, achieving an overall accuracy of 82.0\%. Among the misclassified $H_0$ samples, assignments to $H_1$ are most frequent, indicating ambiguity between the target-free condition and specular returns. Confusion between $H_1$ and $H_2$ suggests that their features can overlap when structured components dominate. Among the misclassified $H_3$ samples, $H_2$ is the most frequent assignment. This pattern is consistent with overlapping echo characteristics between structured multipath and diffuse scattering, although classification errors alone do not establish their cause.

\begin{figure*}[!b]
\centering
\subfloat[]{\includegraphics[width=0.45\textwidth]{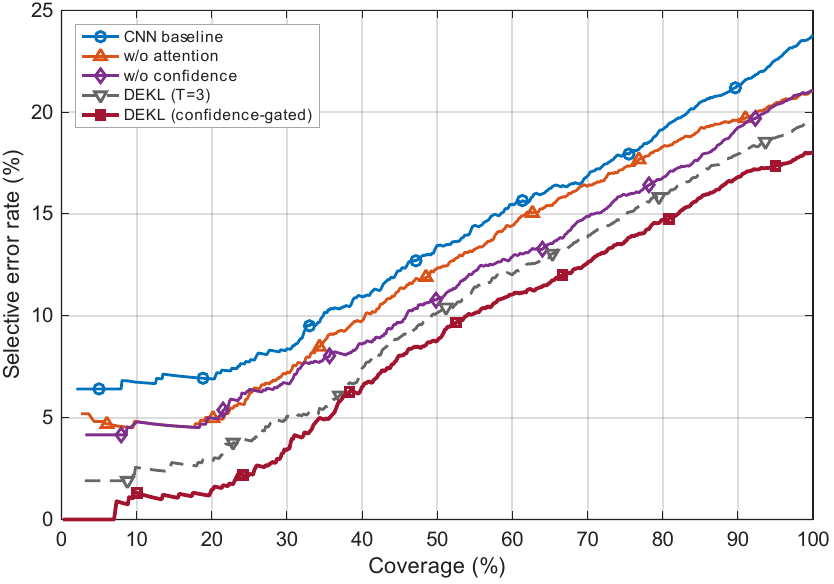}
\label{fig:ablation_risk}}
\hfil
\subfloat[]{\includegraphics[width=0.45\textwidth]{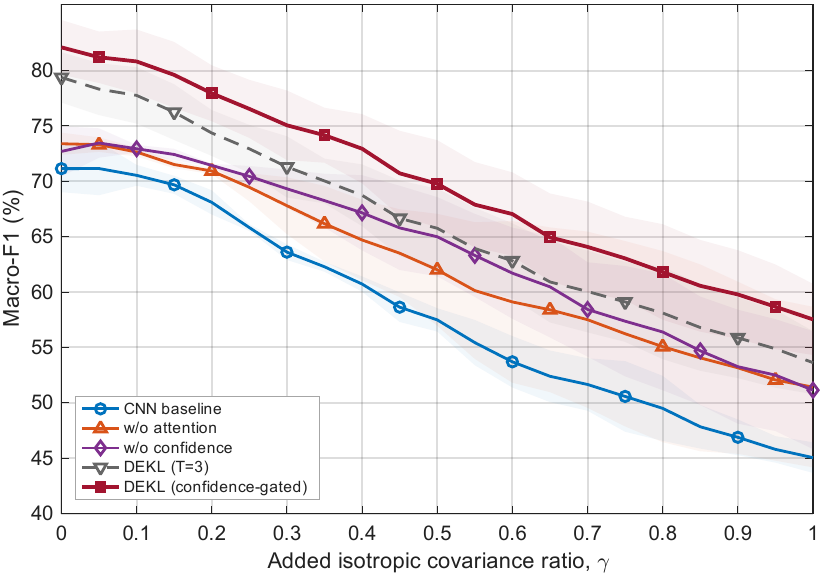}
\label{fig:ablation_covariance}}
\caption{Ablation of DEKL components. (a) Risk--coverage performance of DEKL variants. (b) Macro-F1 under isotropic covariance perturbations. All variants use three decoder steps except DEKL (confidence-gated), which shares the DEKL ($T=3$) weights. In (b), curves and shaded bands show the mean and $\pm1$ sample standard deviation across training seeds 11, 22, and 33 on the same test set.}
\label{fig13}
\end{figure*}

Fig. \ref{fig12} compares classification performance. The specular $H_1$ model has a concentrated array response, whereas $H_2$ contains multiple components that complicate discrimination through simple statistical thresholds. For $H_3$, deviations of measured echoes from the assumed Gaussian model may affect likelihood-based decisions; the comparison does not establish non-Gaussianity. KLIC performs more consistently across classes than the other conventional criteria. DEKL uses covariance-domain representations to discriminate the four hypotheses and achieves the strongest performance in this comparison. These results support its use for the evaluated scattering conditions, while generalization requires measurements covering additional surfaces and geometries.

\subsection{Ablation Study}

To assess the contributions of MOS attention, confidence supervision, and adaptive decoding, five configurations are evaluated on the same 200 measured test samples used in Figs.~\ref{fig11} and~\ref{fig12}. The CNN baseline retains the matched encoder and classification decoder while removing attention and the confidence branch. The w/o attention variant replaces attention with an identity mapping, whereas w/o confidence removes the sigmoid confidence head and its auxiliary loss. These variants and the complete DEKL ($T=3$) model use three decoder steps. DEKL (confidence-gated) shares the complete model's weights within each training seed and selects the second or third output according to Algorithm~\ref{alg:dekl_inference}.

Risk--coverage curves characterize prediction reliability: coverage is the retained sample fraction, and risk is the error rate within that subset. Samples are ranked by the confidence output, or by classification probability for variants without a confidence branch. This offline evaluation retains the original inference rule. Robustness is evaluated using macro-F1, the equally weighted average of the four class-wise F1 scores. Isotropic perturbations are added to the full covariance before feature extraction:
\begin{equation}
\boldsymbol{S}_{\gamma}=\boldsymbol{S}
+\gamma\frac{\operatorname{tr}(\boldsymbol{S})}{n}\boldsymbol{I}_{n},
\qquad \gamma\in[0,1],
\label{eq:47}
\end{equation}
where $n=M_TM_R$ and $\gamma$ controls the added isotropic covariance power relative to the original total power.

Fig. \ref{fig13}(a) shows that confidence-gated DEKL maintains the lowest selective error across the displayed coverage range. The CNN baseline exhibits the highest risk, while removing either attention or the confidence branch degrades performance relative to the complete fixed-step model. These comparisons confirm the contributions of MOS feature weighting and confidence-assisted learning to reliable scattering classification. Gated decoding further improves upon the same-weight fixed-step model by preserving reliable second-step predictions that additional refinement could otherwise degrade. At full coverage, its 18.0\% error corresponds to the 82.0\% overall accuracy in Figs.~\ref{fig11} and~\ref{fig12}, linking the selective-prediction advantage to the measured classification results.

Fig. \ref{fig13}(b) shows a general decrease in macro-F1 as covariance perturbations increase, with confidence-gated DEKL remaining the best-performing configuration throughout the tested range. The complete fixed-step model also outperforms both single-module variants, supporting the complementary contributions of attention and confidence supervision under perturbed inputs. Gated decoding provides a further gain through its adaptive selection of intermediate predictions. At the strongest tested perturbation, DEKL retains a macro-F1 of approximately 57.5\%, compared with 45.0\% for the CNN baseline. Together with its unperturbed performance, which agrees with Fig.~\ref{fig11} at the reported precision, these results demonstrate the classification advantage of DEKL under the investigated covariance conditions.

Using the receive-beamspace input defined in \eqref{eq:42}, single-sample inference requires approximately $1.86\times10^5$ multiply-accumulate operations (MACs) and about 30.8 KB FP32 weight memory. In MATLAB (batch size 1, Ultra 5 225 CPU at 3.30 GHz, 16 GB RAM), offline training averages 300 s and online inference latency averages 2 ms per sample, including feature normalization and network prediction but excluding raw ADC parsing. This efficient inference provides observation-driven scattering model selection for radio-environment representations in wireless-network digital twins.

\FloatBarrier

\section{CONCLUSION}

This paper presented an adaptive framework to identify dominant relay-surface scattering regimes in NLOS millimeter-wave wireless sensing. Candidate echo models and a KLIC-based decision rule establish the statistical formulation, while DEKL uses KLNet for covariance-domain classification and confidence prediction. Analytical simulations, independent FDTD simulations, and radar measurements show improved discrimination over the evaluated information-criterion-based rules. Ablations show contributions of attention, confidence supervision, and gated decoding to selective prediction and covariance-perturbation robustness. The framework links observed echo structure to representative propagation models without direct surface-roughness measurement, providing an observation-driven sensing component for radio-environment and wireless-network digital twins. Under the assumed geometry, the selected category characterizes the dominant received-signal structure, while overlapping echo characteristics can lead to confusion between neighboring regimes. Model selection provides a basis for subsequent NLOS localization and imaging tasks. Future work will examine more complex relay conditions, multitarget scenes, and joint scattering classification and target localization or imaging.

\balance
\bibliographystyle{IEEEtran}
\bibliography{IEEEabrv,REFERENCES}

\begin{thebibliography}{10}
\providecommand{\url}[1]{#1}
\csname url@samestyle\endcsname
\providecommand{\newblock}{\relax}
\providecommand{\bibinfo}[2]{#2}
\providecommand{\BIBentrySTDinterwordspacing}{\spaceskip=0pt\relax}
\providecommand{\BIBentryALTinterwordstretchfactor}{4}
\providecommand{\BIBentryALTinterwordspacing}{\spaceskip=\fontdimen2\font plus
\BIBentryALTinterwordstretchfactor\fontdimen3\font minus \fontdimen4\font\relax}
\providecommand{\BIBforeignlanguage}[2]{{%
\expandafter\ifx\csname l@#1\endcsname\relax
\typeout{** WARNING: IEEEtran.bst: No hyphenation pattern has been}%
\typeout{** loaded for the language `#1'. Using the pattern for}%
\typeout{** the default language instead.}%
\else
\language=\csname l@#1\endcsname
\fi
#2}}
\providecommand{\BIBdecl}{\relax}
\BIBdecl

\bibitem{cao2024raypronet}
G.~Cao and Z.~Peng, ``{RayProNet}: A neural point field framework for radio propagation modeling in {3D} environments,'' \emph{{IEEE} J. Multiscale and Multiphys. Comput. Techn.}, vol.~9, pp. 330--340, 2024.

\bibitem{bakirtzis2025rigorous}
S.~Bakirtzis, K.~Qiu, J.~Chen, H.~Song, J.~Zhang, and I.~Wassell, ``Rigorous indoor wireless communication system simulations with deep learning-based radio propagation models,'' \emph{{IEEE} J. Multiscale and Multiphys. Comput. Techn.}, vol.~10, pp. 58--68, 2025.

\bibitem{pham2024around}
B.-H. Pham, O.~Rabaste, J.~Bosse, I.~Hinostroza, and T.~Chonavel, ````around-the-corner'' radar: Particle filters for non-line-of-sight target tracking in the presence of ambiguities,'' \emph{{IEEE} Trans. Aerosp. Electron. Syst.}, 2024.

\bibitem{danino2024automatic}
S.~Danino and I.~Bilik, ``Automatic multipath annotation for conventional automotive radar datasets,'' \emph{{IEEE} Sensors J.}, vol.~24, no.~8, pp. 13\,500--13\,517, 2024.

\bibitem{xu2023leveraging}
Y.~Xu, G.~Liu, and T.~Jiang, ``Leveraging rough-relay-surface scattering for non-line-of-sight {mmWave} radar sensing,'' \emph{{IEEE} Internet Things J.}, vol.~11, no.~6, pp. 10\,964--10\,978, 2023.

\bibitem{xu2024double}
Y.~Xu, G.~Liu, X.~Lu, C.~Xie, L.~Xue, and T.~Jiang, ``Double sparse structure-enhanced {mmWave} {NLOS} imaging under multiangle relay surface,'' \emph{{IEEE} Trans. Signal Process.}, 2024.

\bibitem{zheng2024detection}
L.~Zheng, J.~Long, M.~Lops, F.~Liu, X.~Hu, and C.~Zhao, ``Detection of ghost targets for automotive radar in the presence of multipath,'' \emph{{IEEE} Trans. Signal Process.}, vol.~72, pp. 2204--2220, 2024.

\bibitem{levy2023mcrb}
M.~Levy-Israel, I.~Bilik, and J.~Tabrikian, ``{MCRB} on {DOA} estimation for automotive {MIMO} radar in the presence of multipath,'' \emph{{IEEE} Trans. Aerosp. Electron. Syst.}, vol.~59, no.~5, pp. 4831--4843, 2023.

\bibitem{xu2025bayesian}
Y.~Xu, G.~Liu, X.~Lu, L.~Xiao, and T.~Jiang, ``Bayesian compressive sensing for {NLOS} {mmWave} imaging under imprecisely multiangle surfaces,'' \emph{{IEEE} Signal Process. Lett.}, 2025.

\bibitem{milczarek2023automatic}
H.~Milczarek, I.~Djurovi{\'c}, C.~Le{\'s}nik, and J.~Jakubowski, ``Automatic classification of frequency-modulated radar waveforms under multipath conditions,'' \emph{{IEEE} Sensors J.}, vol.~23, no.~16, pp. 18\,349--18\,361, 2023.

\bibitem{zhu2023non}
Z.~Zhu, S.~Guo, J.~Chen, S.~Xue, Z.~Xu, P.~Wu, G.~Cui, and L.~Kong, ``Non-line-of-sight targets localization algorithm via joint estimation of {DoD} and {DoA},'' \emph{{IEEE} Trans. Instrum. Meas.}, vol.~72, pp. 1--11, 2023.

\bibitem{borras1998decision}
J.~Borras, P.~Hatrack, and N.~B. Mandayam, ``Decision theoretic framework for {NLOS} identification,'' in \emph{VTC'98. 48th IEEE Vehicular Technology Conference. Pathway to Global Wireless Revolution (Cat. No. 98CH36151)}, vol.~2.\hskip 1em plus 0.5em minus 0.4em\relax IEEE, 1998, pp. 1583--1587.

\bibitem{buttafava2015non}
M.~Buttafava, J.~Zeman, A.~Tosi, K.~Eliceiri, and A.~Velten, ``Non-line-of-sight imaging using a time-gated single photon avalanche diode,'' \emph{Optics express}, vol.~23, no.~16, pp. 20\,997--21\,011, 2015.

\bibitem{fan2024specular}
Z.~Fan, J.~Guo, Y.~Wang, T.~Xiao, H.~Zhang, C.~Zhou, Z.~Chen, P.~Hong, Y.~Guo, and L.-Q. Yan, ``Specular polynomials,'' \emph{ACM Transactions on Graphics (TOG)}, vol.~43, no.~4, pp. 1--13, 2024.

\bibitem{rong2020diffuse}
Y.~Rong, A.~Aubry, A.~De~Maio, and M.~Tang, ``Diffuse multipath exploitation for adaptive detection of range distributed targets,'' \emph{{IEEE} Trans. Signal Process.}, vol.~68, pp. 1197--1212, 2020.

\bibitem{jiang2022survey}
S.~Jiang, W.~Wang, Y.~Miao, W.~Fan, and A.~F. Molisch, ``A survey of dense multipath and its impact on wireless systems,'' \emph{{IEEE} Open J. Antennas Propag.}, vol.~3, pp. 435--460, 2022.

\bibitem{aubry2015diffuse}
A.~Aubry, A.~De~Maio, G.~Foglia, and D.~Orlando, ``Diffuse multipath exploitation for adaptive radar detection,'' \emph{{IEEE} Trans. Signal Process.}, vol.~63, no.~5, pp. 1268--1281, 2015.

\bibitem{wei2021rough}
Y.-W. Wei, C.-F. Wang, C.~Y. Kee, and T.-T. Chia, ``An accurate model for the efficient simulation of electromagnetic scattering from an object above a rough surface with infinite extent,'' \emph{{IEEE} Trans. Antennas Propag.}, vol.~69, no.~2, pp. 1040--1051, 2021.

\bibitem{lian2024distributed}
H.~Lian, M.~Yang, D.~Zhou, and M.~Liu, ``Distributed {MIMO} radar target detection in multipath clutter environments with time reversal,'' \emph{{IEEE} Trans. Aerosp. Electron. Syst.}, 2024.

\bibitem{alburadi2025rapid}
A.~Alburadi, A.~V. Muppala, A.~Y. Nashashibi, H.~N. Shaman, and K.~Sarabandi, ``Rapid indoor mapping and non-line-of-sight imaging using a 228-{GHz} {FMCW} polarimetric radar system,'' \emph{{IEEE} Trans. Geosci. Remote Sens.}, vol.~63, pp. 1--9, 2025.

\bibitem{wen2019joint}
F.~Wen, Z.~Zhang, and G.~Zhang, ``Joint {DOD} and {DOA} estimation for bistatic {MIMO} radar: A covariance trilinear decomposition perspective,'' \emph{{IEEE} Access}, vol.~7, pp. 53\,273--53\,283, 2019.

\bibitem{luo2023effective}
H.~Luo, Z.~Zhu, M.~Jiang, S.~Guo, and G.~Cui, ``An effective multipath ghost recognition method for sparse {MIMO} radar,'' \emph{{IEEE} Trans. Geosci. Remote Sens.}, vol.~61, pp. 1--11, 2023.

\bibitem{guzman2023wave}
M.~F.~D. Guzman and K.~Haneda, ``Analysis of wave-interacting objects in indoor and outdoor environments at 142 {GHz},'' \emph{{IEEE} Trans. Antennas Propag.}, vol.~71, no.~12, pp. 9838--9848, 2023.

\bibitem{chen2025joint}
J.~Chen, X.~Yang, C.~Qiu, Z.~Zhu, P.~Wu, Z.~Xu, S.~Guo, and G.~Cui, ``Joint localization of {LOS} and {NLOS} targets with clutter mitigation via multipath exploitation radar,'' \emph{{IEEE} Trans. Radar Syst.}, 2025.

\bibitem{cao2019sparse}
Z.~Cao, L.~Zhou, and J.~Dai, ``Sparse bayesian approach for {DOD} and {DOA} estimation with bistatic {MIMO} radar,'' \emph{{IEEE} Access}, vol.~7, pp. 155\,335--155\,346, 2019.

\bibitem{addabbo2021adaptive2}
P.~Addabbo, S.~Han, F.~Biondi, G.~Giunta, and D.~Orlando, ``Adaptive radar detection in the presence of multiple alternative hypotheses using {Kullback}--{Leibler} information criterion-part ii: Applications,'' \emph{{IEEE} Trans. Signal Process.}, vol.~69, pp. 3742--3754, 2021.

\bibitem{ge20205g}
Y.~Ge, F.~Wen, H.~Kim, M.~Zhu, F.~Jiang, S.~Kim, L.~Svensson, and H.~Wymeersch, ``{5G} {SLAM} using the clustering and assignment approach with diffuse multipath,'' \emph{Sensors}, vol.~20, no.~16, p. 4656, 2020.

\bibitem{carotenuto2017model}
V.~Carotenuto, A.~De~Maio, D.~Orlando, and P.~Stoica, ``Model order selection rules for covariance structure classification in radar,'' \emph{{IEEE} Trans. Signal Process.}, vol.~65, no.~20, pp. 5305--5317, 2017.

\bibitem{walenczykowska2022application}
M.~Walenczykowska and A.~Kawalec, ``Application of continuous wavelet transform and artificial naural network for automatic radar signal recognition,'' \emph{Sensors}, vol.~22, no.~19, p. 7434, 2022.

\bibitem{ben2008surface}
K.~Ben~Khadhra, ``Surface parameter estimation using bistatic polarimetric x-band measurements,'' Ph.D. dissertation, TU Chemnitz, 2008.

\bibitem{nazari2022random}
M.~E. Nazari and W.~Huang, ``{EM} wave scattering by random surfaces with different contrast and large roughness heights,'' \emph{{IEEE} J. Multiscale and Multiphys. Comput. Techn.}, vol.~7, pp. 252--267, 2022.

\bibitem{rafique2025rough}
S.~Rafique and H.~Arslan, ``A rough surface-aided non-line-of-sight joint sensing and communication system for {THz} band,'' \emph{{IEEE} Open J. Commun. Soc.}, 2025.

\bibitem{ma2019terahertz}
J.~Ma, R.~Shrestha, W.~Zhang, L.~Moeller, and D.~M. Mittleman, ``Terahertz wireless links using diffuse scattering from rough surfaces,'' \emph{{IEEE} Trans. {THz} Sci. Technol.}, vol.~9, no.~5, pp. 463--470, 2019.

\bibitem{barthelme2020model}
A.~Barthelme, R.~Wiesmayr, and W.~Utschick, ``Model order selection in {DoA} scenarios via cross-entropy based machine learning techniques,'' in \emph{ICASSP 2020-2020 IEEE International Conference on Acoustics, Speech and Signal Processing (ICASSP)}.\hskip 1em plus 0.5em minus 0.4em\relax IEEE, 2020, pp. 4622--4626.

\end{thebibliography}

\end{document}